\documentclass[12pt]{article}
\usepackage{amsmath}
\usepackage{graphicx}

\def\papertitlepage{\baselineskip 3.5ex\thispagestyle{empty}}
\def\preprinumber#1#2{\hfill\begin{minipage}{4.2cm} #1
        \par\noindent #2 \end{minipage}}
\allowdisplaybreaks[3]

\begin{document}

\papertitlepage
\setcounter{page}{0}
\preprinumber{KEK-TH-1430}{}
\baselineskip 0.8cm
\vspace*{2.0cm}

\begin{center}
{\Large\bf Non-linear sigma model in de Sitter space}
\end{center}

\begin{center}

Hiroyuki K{\sc itamoto}$^{2)}$
\footnote{E-mail address: kitamoto@post.kek.jp}
and
Yoshihisa K{\sc itazawa}$^{1),2)}$
\footnote{E-mail address: kitazawa@post.kek.jp}\\
\vspace{5mm}
$^{1)}$
{\it KEK Theory Center}\\
{\it Tsukuba, Ibaraki 305-0801, Japan}\\
$^{2)}$
{\it The Graduate University for Advanced Studies (Sokendai)}\\
{\it Department of Particle and Nuclear Physics}\\
{\it Tsukuba, Ibaraki 305-0801, Japan}\\

\end{center}

\vskip 5ex
\baselineskip = 3.5 ex

\begin{center}{\bf Abstract}\end{center}
We investigate infra-red dynamics of the non-linear sigma model in de Sitter space. 
In the presence of minimally coupled massless scalar fields, 
the de Sitter symmetry is dynamically broken and physical constants become time dependent. 
We find that the 
effective coupling constant of the non-linear sigma model becomes time dependent. 
The power counting arguments of the infra-red logarithms indicate that the 
effective cosmological constant also acquires time dependence. 
We find that such infra-red logarithms cancel out in a perturbative investigation up to the two loop level. 
We further demonstrate a non-perturbative non-renormalization of the cosmological constant 
in the large $N$ limit of the non-linear sigma model. 

\vspace*{\fill}
\noindent
December 2010

\newpage
\section{Introduction}
\setcounter{equation}{0}

Cosmic inflation at the early universe and dark energy at the present universe, 
the past and current exponential expansions of the universe are likely to be driven 
by the effective cosmological constant.
The resolution of the cosmological constant problem must address apparent time dependence
of the effective cosmological constant in addition to why it is so small in comparison to the Planck scale.
Our understanding on the history of the effective cosmological
constant has been progressing at a remarkable pace thanks to
increasing more accurate observations on the cosmic micro wave background and  dark energy.
In contrast, the theoretical understanding of the cosmological constant is modest at best.

In fact field theories in de Sitter space are still poorly understood.
As the metric is time dependent, we need to employ Schwinger-Keldysh formalism
to investigate the effects of the interaction \cite{Schwinger1961, Keldysh1964}.
Non-equilibrium physics may play an important role in this problem \cite{Polyakov}.
The constant shift of the cosmic time can be compensated by rescaling the spatial coordinate
to leave the metric invariant. The important issue is whether there is a mechanism to break 
this de Sitter symmetry. 
The local physics probed by the Boltzmann equation respects de Sitter symmetry 
since the local degrees of freedom in field theory are time independent \cite{KK}.

On the other hand,
the degrees of freedom outside the cosmological horizon increase with cosmic evolution. 
This increase gives rise to a growing time dependence 
to the propagator of a massless and minimally coupled scalar field and gravitational field. 
In fact it is a direct consequence of their scale invariant fluctuation spectrum.
In some field theoretic models on de Sitter space, the de Sitter symmetry is dynamically broken and the 
effective cosmological constant acquires a time dependence 
through such an effect. 
The relevance of this infra-red effect to the cosmological constant problem has been pointed out in \cite{Woodard1996}. 

In $\varphi^4$ theory, the infra-red effects to the cosmological constant has been investigated in perturbation theory 
\cite{Woodard2002}. 
In generic models, the maximum power of the infra-red logarithms can be estimated at each order in 
Schwinger-Keldysh perturbation theory
\cite{Weinberg}.
These results indicate that the perturbation theory breaks down 
if the de Sitter expansion continues long enough. 
In such a situation,
we have to investigate the infra-red effect non-perturbatively. 
Remarkably the leading non-perturbative infra-red effect can be evaluated 
by the stochastic approach \cite{Starobinsky1994,Woodard2005}. 
However in a general model with derivative interactions, 
we don't know how to evaluate the non-perturbative infra-red effect. 

As a model with derivative interactions, 
we investigate the non-linear sigma model in this paper. 
The global symmetry guarantees that the non-linear sigma model contains massless and minimally coupled scalar fields 
while a fine tuning is necessary in the case of polynomial interactions. 
Furthermore we can perform some non-perturbative investigations 
as it is exactly solvable in the large $N$ limit. 
Another point is that there is some similarity to the Einstein action as it consists of the derivative interactions of 
the metric tensor field.

The organization of this paper is as follows. 
In Section $2$, we recall a scalar field theory in de Sitter space, in particular its infra-red behavior. 
In Section $3$, we investigate the quantum expectation value of the energy-momentum tensor
in an interacting scalar field theory through a polynomial potential.
In Section $4$, we investigate the energy-momentum tensor of the non-linear sigma model. 
We find unexpected cancellations of the IR logarithms beyond the IR power counting arguments
both in a concrete perturbation theory and in a non-perturbative investigation.
These are our main results in this paper. 
In Section $5$, we evaluate the trace of the energy momentum tensor to confirm the results in Section $3$ and $4$. 
We conclude with discussions in Section $6$. 

\section{Scalar field in de Sitter space}
\setcounter{equation}{0}
We introduce a massless scalar field theory in de Sitter(dS) space in this section. 
In particular, we focus on its infra-red(IR) dynamics which makes physical quantities time dependent. 

In the Poincar\'{e} coordinate, the metric in dS space is
\begin{equation}
ds^2=-dt^2+a^2(t)d{\bf x}^2,\hspace{1em}a(t)=e^{Ht}, 
\end{equation}
where the dimension of dS space is taken as $D=4$ and $H$ is the Hubble constant. 
In the conformally flat coordinate,
\begin{equation}
g_{\mu\nu}=a^2(\tau)\eta_{\mu\nu},\hspace{1em}a(\tau)=-\frac{1}{H\tau}. 
\end{equation}
Here the conformal time $\tau$ is related to the cosmic time $t$ as $\tau\equiv-\frac{1}{H}e^{-Ht}$. 

The quadratic action for a massless scalar field which is minimally coupled to the dS background is
\begin{equation}\begin{split}
S_{matter}=\frac{1}{2}\int\sqrt{-g}d^4x\ [-g^{\mu\nu}\partial_\mu\varphi\partial_\nu\varphi].
\end{split}\end{equation}
The positive frequency solution of the equation of motion with respect to this action is
\begin{equation}
\phi_{{\bf p}}(x)=\frac{H\tau}{\sqrt{2p}}(1-i\frac{1}{p\tau})\ e^{-ip\tau+i{\bf p}\cdot{\bf x}},
\end{equation}
where $p=|\bf{p}|$. 
We expand the scalar field as
\begin{equation}
\varphi (x) = \int \frac{d^3p}{(2\pi)^3}\left( a_{{\bf p}}\phi_{{\bf p}}(x)
+ a_{{\bf p}}^{\dagger}\phi_{{\bf p}}^*(x)\right). 
\end{equation}
We consider the Bunch-Davies vacuum $|0\rangle$ which is annihilated by all the destruction operators 
$\forall a_{{\bf p}}|0\rangle=0$.
The propagator in such a vacuum is 
\begin{equation}\begin{split}
\langle\varphi(x)\varphi(x')\rangle
=&\int \frac{d^3p}{(2\pi)^3}\ \phi_{{\bf p}}(x)\phi_{{\bf p}}^*(x')\\
=&\int \frac{d^3p}{(2\pi)^3}\ \frac{H^2\tau\tau'}{2p}(1-i\frac{1}{p\tau})(1+i\frac{1}{p\tau'})
\ e^{-ip(\tau-\tau')+i{\bf p}\cdot({\bf x}-{\bf x}')}.
\end{split}\end{equation}

Let us estimate the magnitude of the quantum fluctuation by taking the coincident limit of the propagator.
It consists of the contributions from inside and outside the cosmological horizon as follows 
\begin{equation}
\langle\varphi(x)\varphi(x)\rangle\sim \int_{P>H}\frac{d^3P}{(2\pi)^3}\frac{1}{2P}+H^2\int_{P<H}\frac{d^3P}{(2\pi)^3}\frac{1}{2P^3}, 
\end{equation}
where $P$ denotes the physical momentum $P\equiv p/e^{Ht}=H|\tau|p$.
The UV contribution $(P>H)$ is quadratically divergent just like in Minkowski space. 
It can be regularized and renormalized in an identical way.
The logarithmic IR divergence due to the contributions from outside the cosmological horizon $(P<H)$ is specific to de Sitter space.
To regularize this IR divergence, we introduce an IR cut-off $\varepsilon_0$ : 
\begin{equation}
\int^H_{\varepsilon_0H|\tau|}dP. 
\end{equation}
Here $\varepsilon_0$ fixes the minimum value of the comoving momentum \cite{Lyth2007}.

With this prescription, more degrees of freedom go out of the cosmological horizon $P=H$ with cosmic evolution. 
In contrast, the ultra-violet(UV) cut-off $\Lambda_{UV}$ fixes the maximum value of the physical momentum
\begin{equation}
\int^{\Lambda_{UV}}_HdP. 
\end{equation}
Therefore, while the degrees of freedom inside the cosmological horizon remains constant, 
the degrees of freedom outside the cosmological horizon increases as time goes on 
\begin{equation}
\int dP=\underset{const}{\underline{\int^{\Lambda_{UV}}_HdP}}+\underset{increase}{\underline{\int^H_{\varepsilon_0H|\tau|}dP}}. 
\end{equation}
The contribution from outside the cosmological horizon gives a growing time dependence to the propagator 
\begin{equation}\begin{split}
\langle\varphi(x)\varphi(x)\rangle&=(UV\ const)+\frac{H^2}{4\pi^2}\int^H_{\varepsilon_0H|\tau|}\frac{dP}{P}\\
&=(UV\ const)+\frac{H^2}{4\pi^2}\big|\log(\varepsilon_0|\tau|)\big|.
\end{split}\end{equation}
Physically speaking, we consider a situation that a universe with a finite 
spatial extension or a finite region of space starts de Sitter expansion at an initial time $t_i$.
The IR cut-off $\varepsilon_0$ is identified with the initial time $t_i$ as 
\begin{equation}
\big|\log(\varepsilon_0|\tau|)\big|=\log e^{H(t-t_i)},
\hspace{1em}t_i\equiv\frac{1}{H}\log\frac{\varepsilon_0}{H}. 
\end{equation}
Henceforth we adopt the following setting for simplicity
\begin{equation}
\varepsilon_0=H\Leftrightarrow t_i=0. 
\end{equation}
In this setting, the propagator is 
\begin{equation}
\langle\varphi(x)\varphi(x)\rangle=(UV\ const)+\frac{H^2}{4\pi^2}\log a(\tau). 
\end{equation}

We have thus recapitulated the characteristic feature of the propagator of a minimally coupled 
massless scalar field in de Sitter space \cite{Vilenkin1982,Linde1982,Allen1985}. 
The metric of de Sitter space is invariant under the time translation $t\rightarrow t+c$ if we simultaneously rescale the spatial coordinate ${\bf x}\rightarrow e^{-Hc}{\bf x}$. 
This de Sitter invariance is not respected
by the propagator of a minimally coupled massless scalar field due to IR divergences. 
The time dependence of the propagator induces time dependences to physical quantities
if we consider the effects of the interaction. 
We investigate such effects which cause the dynamical break down of de Sitter invariance in the subsequent sections. 

\section{Energy-momentum tensor}
\setcounter{equation}{0}

In this section, we investigate the energy-momentum tensor $T_{\mu\nu}$ for a scalar field
which appears on the right-hand side of the Einstein equation.
\begin{equation}\begin{split}
R_{\mu\nu}-\frac{1}{2}g_{\mu\nu}R+\Lambda g_{\mu\nu}&=\kappa T_{\mu\nu},\hspace{1em}\kappa=8\pi G,\\
T_{\mu\nu}&\equiv\frac{-2}{\sqrt{-g}}\frac{\delta S_{matter}}{\delta g^{\mu\nu}},
\end{split}\label{Eeq}\end{equation}
where $\Lambda$ is the cosmological constant and $G$ is the Newton's constant. 
The de Sitter invariance implies that the quantum expectation value of $T_{\mu\nu}$ is proportional to
the metric tensor $g_{\mu\nu}$. It thus contributes to the effective cosmological constant.

\subsection{Contribution from a free field}

In the minimally coupled massless free scalar field theory, 
the vacuum expectation value(vev) of the energy-momentum tensor is
\begin{equation}
\langle T_{\mu\nu}\rangle=(\delta_\mu^{\ \rho}\delta_\nu^{\ \sigma}-\frac{1}{2}\eta_{\mu\nu}\eta^{\rho\sigma})
\langle\partial_\rho\varphi\partial_\sigma\varphi\rangle. 
\end{equation}
In order to evaluate the quantum expectation value, 
we need to regularize UV and IR divergences of the theory. 
We adopt the dimensional regularization with $D=4-\varepsilon$ as a regularization for UV divergences. 
Note that the de Sitter invariance of the propagator is respected in this regularization. 
It is because the UV divergences are represented by $1/\varepsilon$, that is, a constant in the dimensional regularization.
We regularize the IR divergences by considering a universe with a finite 
spatial extension as we explained in the preceding section.

In this way, 
the propagator for a massless and minimally coupled scalar field is constructed as follows \cite{Miao}
\begin{equation}
\langle\varphi(x)\varphi(x')\rangle=\alpha\{\gamma(y)+\beta\log(a(\tau)a(\tau'))\}, 
\label{G}\end{equation}
\begin{equation}
\alpha\equiv\left(\frac{H}{2\pi}\right)^2\left(\frac{H}{\sqrt{\pi}}\right)^{-\varepsilon}\Gamma(2-\frac{\varepsilon}{2}),
\hspace{1em}\beta\equiv\frac{1}{4^{1-\frac{\varepsilon}{2}}}\frac{\Gamma(3-\varepsilon)}{\Gamma^2(2-\frac{\varepsilon}{2})},
\label{alpha}\end{equation}
\begin{equation}\begin{split}
\gamma(y)\equiv&\ \frac{1}{1-\frac{\varepsilon}{2}}\frac{1}{y^{1-\frac{\varepsilon}{2}}}
+\frac{1}{4^{1-\frac{\varepsilon}{2}}}\frac{\Gamma(3-\varepsilon)}{\Gamma^2(2-\frac{\varepsilon}{2})}
\ \delta
-(1-\frac{\varepsilon}{4})\frac{y^\frac{\varepsilon}{2}}{\varepsilon}\\
&+\sum^{\infty}_{n=1}\left[\frac{1}{n}\frac{\Gamma(3+n-\varepsilon)}{\Gamma(2+n-\frac{\varepsilon}{2})\Gamma(2-\frac{\varepsilon}{2})}4^\frac{\varepsilon}{2}
-\frac{1}{n+\frac{\varepsilon}{2}}\frac{\Gamma(3+n-\frac{\varepsilon}{2})}{(n+1)!\Gamma(2-\frac{\varepsilon}{2})}y^\frac{\varepsilon}{2}\right]\frac{y^n}{4^{n+1}},\\
\delta\equiv&-\psi(-1+\frac{\varepsilon}{2})+\psi(\frac{3}{2}-\frac{\varepsilon}{2})+\psi(3-\varepsilon)+\psi(1), 
\end{split}\label{gamma}\end{equation}
where $y$ is a dS invariant distance function
\begin{equation}
y\equiv\frac{-(\tau-\tau')^2+({\bf x}-{\bf x}')^2}{\tau\tau'}. 
\end{equation}
Note that the $\log(a(\tau)a(\tau'))$ term comes from the IR cut-off and breaks the dS invariance.
If a field is massive, the IR cut-off is not required and the dS invariance is retained
\cite{Marolf, Hollands}.
However there is no smooth massless limit and we need to cope with large IR quantum effects anyway.
If we consider a universe with a finite spatial extension, 
a massive field behaves just like a massless field 
until when the mass term cuts off the IR growth of the propagator at $t\sim H/m^2$
\cite{Vilenkin1982, Linde1982, Starobinsky1982, Finelli, Smit}.
We are interested in the coincident limit of the propagator and its derivatives when we evaluate 
the quantum expectation value of the energy-momentum tensor. Although 
it looks cumbersome that $\gamma(y)$ has infinite terms,
only finite terms contribute in our investigation. 

In the free field theory, the following two derivative operation cancels out the IR logarithm 
\begin{equation}
\partial_\rho\partial_\sigma'\log \big(a(\tau)a(\tau')\big)=0. 
\end{equation}
Only the term which is proportional to $y$ contributes to the vev of the energy-momentum tensor
\begin{equation}
\lim_{x'\to x}\partial_\rho\partial_\sigma'y=-2H^2g_{\rho\sigma}, 
\end{equation}
\begin{equation}
\langle T_{\mu\nu}\rangle=\frac{3H^4}{32\pi^2}g_{\mu\nu}. 
\label{freeT}\end{equation}
As a result, the vev of the energy-momentum tensor is proportional to $g_{\mu\nu}$ in the free field theory. 
From the Einstein equation (\ref{Eeq}), the effective cosmological constant is 
\begin{equation}
\Lambda_{eff}=\Lambda-\kappa\frac{3H^4}{32\pi^2}. 
\end{equation}
The effective cosmological constant has no time dependence in the free field theory. 
It is because there is no IR divergences when we evaluate the quantum expectation value
of the energy-momentum tensor in a free scalar field theory. 

In this paper, we work with the Poincar\'{e} coordinate. 
The propagator and 
the energy-momentum tensor for a free field is investigated by using the global coordinate in \cite{Allen1987,Folacci}. 
The result is a little different from (\ref{freeT}). 
However the difference rapidly vanishes at late times with the spatial expansion. 
To obtain the growing time dependence, we need to consider the interaction effects.
It is necessary that there is a propagator left intact by the derivatives 
$\langle\varphi(x)\varphi(x')\rangle$ in the relevant diagram. 

Before investigating interacting field theory, we refer to the conformal anomaly. 
The conformal anomaly also contributes to the vev of the energy-momentum tensor \cite{BD}.
In the case of the minimally coupled scalar field in de Sitter space, it leads to the
following energy momentum tensor in addition
\begin{equation}
\langle T_{\mu\nu}\rangle=\frac{29H^4}{15\cdot 64\pi^2}g_{\mu\nu}. 
\end{equation}
This contribution has no time dependence as the conformal anomaly is the UV effect. 
In this paper, we focus on the IR contribution to the cosmological constant 
and its possible time dependence. 

\subsection{Contribution from an interacting field : Perturbative effect}

Here we investigate the perturbative effects to the vev of the energy-momentum tensor in an interacting 
field theory. 

First, we consider a scalar field theory with an interaction potential  
\begin{equation}
S_{matter}=\int\sqrt{-g}d^4x\ [-\frac{1}{2}g^{\mu\nu}\partial_\mu\varphi\partial_\nu\varphi-V(\varphi)]. 
\label{Vaction}\end{equation}
In this case, the energy-momentum tensor is given as follows 
\begin{equation}
T_{\mu\nu}=\ (\delta_\mu^{\ \rho}\delta_\nu^{\ \sigma}-\frac{1}{2}\eta_{\mu\nu}\eta^{\rho\sigma})\partial_\rho\varphi\partial_\sigma\varphi
-g_{\mu\nu}V(\varphi). 
\end{equation}
As we have recalled in the previous section, the propagator of a massless minimally coupled scalar field contains 
the time dependent term $\log(a(\tau)a(\tau'))$.
So the vev of the potential becomes time dependent. $\log(a(\tau))=Ht$ factor grows with cosmic expansion which eventually
gives rise to a large IR quantum effect.
The maximum power of the IR logarithms can be estimated in each order of Schwinger-Keldysh perturbation theory.
In the case of the polynomial interaction, each propagator could produce a single IR log factor.
Although the retarded propagator resulting from the commutator in the Schwinger-Keldysh perturbation theory
does not contain any IR log factor, the associated time integration produces a single IR factor nevertheless.
In the end, the maximum power of the IR logarithms can be estimated by counting the number of the propagators
in a diagram.
 
In this way the leading IR contributions coming from the potential are estimated as
\begin{equation}
-g_{\mu\nu}\langle V(\varphi)\rangle
=-g_{\mu\nu}V(\tau)
\simeq -g_{\mu\nu}\sum_{n=1}V_n\lambda^n\log^{mn}a(\tau), \hspace{1em}m\in{\bf N}. 
\label{potential}\end{equation}
Here we have focused on the dominant effect at late times when $\log a(\tau)\gg 1$. 
At the $n$-th order of the coupling constant in $\lambda \varphi ^{2m}$ theory, 
we retain the term of order $(\lambda \log^m(a))^n$ where the power of the IR logarithm is maximum.
Even if $\lambda$ is small, higher order terms cannot be neglected as $\lambda \log^m(a)$ approaches $\mathcal{O}(1)$. Thus we need to sum these leading logarithms first in order to understand the IR effects non-perturbatively. Sub-leading terms are suppressed by powers of $\lambda$ just like the resummation of
IR logarithms in QCD.
Since the contribution from the potential is proportional to $g_{\mu\nu}$, it gives rise to a time dependent effective cosmological constant. 

The energy-momentum tensor is covariantly conserved under the use of the equation of motion : 
\begin{equation}\begin{split}
0=\int d^4x\ \frac{\delta S_{matter}}{\delta \varphi}\mathcal{L}_\xi \varphi
&=\frac{-1}{2}\int\sqrt{-g}d^4x\ T^{\mu\nu}\mathcal{L}_\xi g_{\mu\nu}\\
&=\int\sqrt{-g}d^4x\ D_\mu T^{\mu\nu}\xi_\nu\\
\Rightarrow \ D_\mu T^\mu_{\ \nu}=0, &
\end{split}\end{equation}
where $\mathcal{L}_\xi$ denotes the Lie derivative. 
If the energy-momentum tensor is proportional to $g_{\mu\nu}$, 
its time dependent coefficient is not consistent with the conservation law. 

The contribution from the kinetic term resolves this puzzle. 
$\langle\partial_\rho\varphi\partial_\sigma\varphi\rangle$ possesses the following structure
\begin{equation}\begin{split}
\langle\partial_\rho\varphi\partial_\sigma\varphi\rangle&=-g_{\rho\sigma}\frac{3H^4}{32\pi^4}
+a^2(\tau)\delta_\rho^{\ 0}\delta_\sigma^{\ 0}K^1(\tau)+g_{\rho\sigma}K^2(\tau),\\
K^1(\tau)&\simeq\sum_{n=1}K^1_n\lambda^n\log^{mn-1}a(\tau),\\
K^2(\tau)&\simeq\sum_{n=1}K^2_n\lambda^n\log^{mn-1}a(\tau). 
\end{split}\label{kinetic}\end{equation}
Here the first term is the the free field theory contribution at the one loop level. 
Note that the kinetic term is sub-dominant in comparison to the potential term except at the one loop level. 
It is because taking two derivative operations weakens the IR effect. 
So in the energy-momentum tensor, 
we can neglect the contribution $g_{\mu\nu}\{\frac{1}{2}K^1(\tau)-K^2(\tau)\}$. 
That is, we estimate the vev of the energy-momentum tensor as follows 
\begin{equation}
\langle T_{\mu\nu}\rangle\simeq g_{\mu\nu}\frac{3H^4}{32\pi^4}+a^2(\tau)\delta_\mu^{\ 0}\delta_\nu^{\ 0}K^1(\tau)-g_{\mu\nu}V(\tau). 
\label{EMT}\end{equation}
In the case that the potential term acquires time dependence, the following identity follows from 
the conservation law 
\begin{equation}
K^1(\tau)=\frac{\tau}{3}\frac{d}{d\tau}V(\tau) \ \Leftrightarrow\ K^1_n(\tau)=-\frac{mn}{3}V_n(\tau). 
\label{conservation}\end{equation}
The de Sitter invariance is thus broken down to the spatial rotation 
and spatial translation invariance due to the IR quantum effect.

We summarize the IR effects to the expectation value of the energy-momentum tensor here. 
At the one loop level, the contribution is identical to that of a free field and there is no IR effects. 
At $\mathcal{O}(\lambda^n)\ (n\geq 1)$ when the effect of the interaction becomes important, 
the potential term becomes dominant at late times in de Sitter expansion. 
The dominant term is proportional to $g_{\mu\nu}$ and so contributes to the effective cosmological constant 
\begin{equation}
\Lambda_{eff}=\Lambda-\kappa\frac{3H^4}{32\pi^4}+\kappa V(\tau). 
\end{equation}
The kinetic term at $\mathcal{O}(\lambda^n)\ (n\geq 1)$ is sub-dominant and 
contains the term which is proportional to $\delta_\mu^{\ 0}\delta_\nu^{\ 0}$. 
Such a term is related to the effective cosmological constant due to the conservation law 
\begin{equation}
D_\mu\langle T^\mu_{\ \nu}\rangle
\simeq \delta_\nu^{\ 0}\big\{\frac{3}{\tau}K^1(\tau)-\frac{d}{d\tau}V(\tau)\big\}=0. 
\end{equation}

We can explicitly check that the condition (\ref{conservation}) is satisfied, 
for example, in $\varphi^4$ theory. 
The 2-loop expectation value in $\varphi^4$ theory is evaluated as follows \cite{Woodard2002}
\begin{equation}
-g_{\mu\nu}\langle V(\varphi)\rangle = -g_{\mu\nu}\frac{\lambda H^4}{2^7\pi^4}\log^2a(\tau), 
\label{4Vp}\end{equation}
\begin{equation}
\langle\partial_\rho\varphi\partial_\sigma\varphi\rangle\simeq -g_{\rho\sigma}\frac{3H^4}{32\pi^4}
-a^2(\tau)\delta_\rho^{\ 0}\delta_\sigma^{\ 0}\frac{\lambda H^4}{2^6\cdot 3\pi^4}\log a(\tau). 
\end{equation}
The effective cosmological constant increases as time goes on
\begin{equation}
\Lambda_{eff}=\Lambda-\kappa\frac{3H^4}{32\pi^4}+\kappa \frac{\lambda H^4}{2^7\pi^4}\log^2a(\tau). 
\label{4pLambda}\end{equation}
Nevertheless the energy-momentum tensor is covariantly conserved
\begin{equation}
D_\mu\langle T^\mu_{\ \nu}\rangle\simeq
\delta_\nu^{\ 0}\big\{-\frac{3}{\tau}\frac{\lambda H^4}{2^6\cdot 3\pi^4}\log a(\tau)-\frac{d}{d\tau}\frac{\lambda H^4}{2^7\pi^4}\log^2a(\tau)\big\}=0. 
\end{equation}

The structure of the vev of the energy-momentum tensor (\ref{EMT}) holds in a generic model.
The coefficients of the two independent tensors $g_{\mu\nu}$ and $\delta_{\mu}^{\ 0}\delta_{\nu}^{\ 0}$
are related as (\ref{conservation}) because of the conservation law. 
In the models we reviewed in this section, the dominant contribution comes from the potential term. 
However in a model with derivative interactions only, we need to estimate the contribution 
from the kinetic term. 
As a model with derivative interactions, 
we investigate the non-linear sigma model in the next section. 

\subsection{Contribution from an interacting field : Non-perturbative effect}

Perturbation theory eventually breaks down when $\lambda\log^m a(\tau)\sim H^{-2m+4}$. 
So we need a tool to investigate the non-perturbative effect in such a regime. 
There is a stochastic approach for investigating such a non-perturbative effect 
\cite{Starobinsky1994,Woodard2005}.  It can be regarded as a resummation procedure of the leading
IR logarithms due to an interaction potential. 
Here we briefly recall this prescription. 

In a minimally coupled scalar field theory with a potential, 
the equation of motion is
\begin{equation}
\ddot{\varphi}+3H\dot{\varphi}-\frac{1}{a^2}\partial_i^2\varphi+V'(\varphi)=0, 
\end{equation} 
where $\dot{\varphi}\equiv \frac{\partial}{\partial t}\varphi$. 
The equation of motion can be integrated as
\begin{equation}
\varphi(x)=\varphi_0(x)-i\int^t_0 dt'a^3(t')\int d^3x'\ G^R(x,x')V'(\varphi(x')), 
\label{EoM1}\end{equation} 
where $\varphi_0(x)$ denotes a free field and $G^R(x,x')$ is the retarded propagator
\begin{equation}
G^R(x,x')=\theta(t-t')[\langle\varphi_0(x)\varphi_0(x')\rangle-\langle\varphi_0(x')\varphi_0(x)\rangle]. 
\end{equation} 
As we are interested in the dominant IR effect at late times, 
we extract the contribution from outside the cosmological horizon 
\begin{equation}
\varphi_0(x)\simeq\int\frac{d^3p}{(2\pi)^3}\ \theta(Ha(t)-p)
\big(a_{{\bf p}}\frac{H}{\sqrt{2p^3}}e^{+i{\bf p}\cdot{\bf x}}+a^\dagger_{{\bf p}}\frac{H}{\sqrt{2p^3}}e^{-i{\bf p}\cdot{\bf x}}\big). 
\label{free}\end{equation} 
For the same reason, we extract the leading IR contribution of the propagator
\begin{equation}\begin{split}
G^R(x,x')&\simeq \theta(t-t')\int\frac{d^3p}{(2\pi)^3}\ \frac{-i}{3H}\left(\frac{1}{a^3(t')}-\frac{1}{a^3(t)}\right)e^{+i{\bf p}\cdot({\bf x}-{\bf x}')}\\
&=\frac{-i}{3H}\left(\frac{1}{a^3(t')}-\frac{1}{a^3(t)}\right)\theta(t-t')\delta^{(3)}({\bf x}-{\bf x'}). 
\label{R}\end{split}\end{equation} 
By substituting (\ref{free}) and (\ref{R}) to Eq.(\ref{EoM1}), 
\begin{equation}
\varphi(x)=\varphi_0(x)-\frac{1}{3H}\int^t_0 dt'\ V'(\varphi(t',{\bf x})), 
\label{EoM2}\end{equation} 
where we have neglected the term : $a^{-3}(t)\int^t_0dt'a^3(t')V(\varphi(t',{\bf x})$ because it is sub-dominant.
 
By differentiating Eq.(\ref{EoM2}) with respect to $t$, we obtain the Langevin equation with the white noise 
\begin{equation}
\dot{\varphi}(x)=\dot{\varphi_0}(x)-\frac{1}{3H}V'(\varphi(x)),\hspace{1em}\langle\dot{\varphi}_0(x)\dot{\varphi}_0(x')\rangle=\frac{H^3}{4\pi^2}\delta(t-t'). 
\label{EoM3}\end{equation} 
It describes a random walk in the field space. Since the fractal dimension of the random walk is two, the propagator
grows linearly with the cosmic time $Ht = \log a$ at the initial stage.
The Langevin equation is equivalent to the Fokker-Planck equation 
\begin{equation}
\dot{\rho}(t,\varphi)=\frac{1}{3H}\frac{\partial}{\partial\varphi}\big[V'(\varphi)\rho(t,\varphi)\big]+\frac{H^3}{8\pi^2}\frac{\partial^2}{\partial\varphi^2}\rho(t,\varphi), 
\label{EoM4}\end{equation} 
where $\rho(t,\varphi)$ is the probability density. The vevs of the operators are given by 
\begin{equation}
\langle\varphi^{2n}(x)\rangle=\int^\infty_{-\infty}d\omega\ \omega^{2n}\rho(t,\omega). 
\label{density}\end{equation} 

We can reproduce the leading log terms in the perturbative expansion in this approach.
Furthermore it allowed us to determine the non-perturbative effect at $t\to\infty$ when 
we assume that an equilibrium state is established at $t\to\infty$ : $\rho(t,\varphi)\to\rho_\infty(\varphi)$. 
In this assumption, the Fokker-Planck equation is
\begin{equation}
0=\frac{1}{3H}V'(\varphi)\rho_\infty(\varphi)+\frac{H^3}{8\pi^2}\frac{\partial}{\partial\varphi}\rho_\infty(\varphi). 
\label{EoM5}\end{equation} 
The solution is
\begin{equation}
\rho_\infty(\varphi)=N\exp\left(-\frac{8\pi^2}{3H^4}V(\varphi)\right), 
\label{densitylimit}\end{equation} 
where $N$ is the normalization factor : $\int^\infty_{-\infty}d\omega\ \rho_\infty(\omega)=1$. 
From (\ref{density}) and (\ref{densitylimit}), we can evaluate the vevs of the operators at $t\to\infty$, especially the vev of the potential. 
The dynamics of the final equilibrium state may be understood by that of the zero mode
in a Euclidean scalar field theory on $S_4$
\cite{Rajaraman2010} .

For example, in $\varphi^4$ theory, the probability density is 
\begin{equation}
\rho_\infty(\varphi)=\frac{2}{\Gamma(\frac{1}{4})}\left(\frac{\pi^2\lambda}{9H^4}\right)^\frac{1}{4}
\exp\left(-\frac{\pi^2}{9}\lambda\frac{\varphi^4}{H^4}\right). 
\end{equation} 
The vev of the potential is
\begin{equation}
\langle V(\varphi)\rangle=\frac{3H^4}{32\pi^2}. 
\label{4Vs}\end{equation} 
As we observe in (\ref{4Vp}), the 2-loop effect increases the vev of the potential as time goes on. 
It is because the magnitude of $\varphi$ field grows due to a random walk in a stochastic approach. 
Eventually the drift force
due to the potential becomes important and reaches an equilibrium.
Thus the following consistent picture emerges, namely the effective cosmological constant increases 
at the initial stage
and the growth is eventually saturated at a constant value.  

To evaluate the vev of the energy-momentum tensor, 
we also need to consider the kinetic term. 
Note that we retain only the leading IR effect in the stochastic approach. 
The kinetic term is sub-dominant in comparison to the potential term except at the one loop level. 
So we can't calculate the kinetic term directly in the stochastic approach. 
Nevertheless the structure of the kinetic term is constrained by the conservation law. 
As the potential term approaches a constant at $t\to\infty$,
the de Sitter symmetry breaking contribution from the kinetic term also vanishes. 
That is, the term which is proportional to $\delta_\mu^{\ 0}\delta_\nu^{\ 0}$ approaches to $0$. 
Of course, it is possible that the sub-leading terms give a finite contribution to the cosmological constant. 
However these contributions are $\mathcal{O}(\lambda^\frac{1}{m})$ at most. 
We can neglect them if $\lambda\ll H^{-2m+4}$. 

From (\ref{4Vs}), the effective cosmological constant at $t\to\infty$ is as follow in $\varphi^4$ theory 
\begin{equation}
\Lambda_{eff}=\Lambda-\kappa\frac{3H^4}{32\pi^2}+\kappa\frac{3H^4}{32\pi^2}=\Lambda. 
\label{4sLambda}\end{equation}
The contribution from the kinetic term at the one loop level and the non-perturbative contribution from the
potential term cancel out each other. 
It is an accident in $\varphi^4$ theory. 
In $\varphi^{2m}\ (m\not=2)$ theory, there remains a finite contribution to the cosmological constant. 

This stochastic approach has been applied to investigate non-perturbative IR effects in
Yukawa theory \cite{Woodard2006} and Scalar QED \cite{Woodard2007}. 
These models reduce to a scalar field theory with a potential (\ref{Vaction}) after integrating out
the conformally coupled scalar fields, Dirac fields or vector fields. 

Another non-perturbative approach to investigate IR effects is to consider the large $N$ limit
where $N$ counts the number of scalar field.
In such a limit, we can solve the model by a saddle point approximation.
The action for $\varphi^4$ theory with $O(N)$ symmetry can be expressed as follows 
\begin{equation}
S_{matter}=\int\sqrt{-g}d^4x \ [-\frac{1}{2}g^{\mu\nu}\partial_\mu\varphi^i\partial_\nu\varphi^i
-\frac{\chi}{2}(\varphi^i)^2+\frac{N}{2\lambda}\chi^2],   
\label{4N}\end{equation}
where $i=1\cdots N$ and $\chi$ is an auxiliary field. 
By differentiating the action with respect to $\chi$, we find that $\chi$ represents a composite
operator
\begin{equation}
(\varphi^i)^2=\frac{2N}{\lambda}\chi. 
\label{4N1}\end{equation}
In the large $N$ limit, we can neglect the fluctuation of $\chi$. 
So the action (\ref{4N}) reduces to a free massive scalar field theory plus the constant term $N\chi^2/2\lambda$. 

Here $\chi$ acts as the mass of scalar fields $m^2=\chi$. 
At the initial stage, $m^2$ grows with time as is determined from (\ref{4N1})
\begin{equation}
m^2 = \lambda {H^2\over 8\pi^2}\log a(\tau ). 
\end{equation}
Eventually it approaches a constant which is self-consistently determined as
\begin{equation}
m^2=\lambda\frac{3H^4}{16\pi^2m^2} , 
\label{4N2}\end{equation}
where we have assumed that $m^2/H^2$ is small.
From (\ref{4N2}), 
\begin{equation}
m^2=\chi=\frac{\sqrt{3\lambda}H^2}{4\pi}. 
\label{4N3}\end{equation}
Recall that the propagator of a massive scalar field has the dS invariance. 
Therefore the vev of the energy-momentum tensor is written as follows 
\begin{equation}
\langle T_{\mu\nu}\rangle=\frac{g_{\mu\nu}}{4}\langle T_\rho^{\ \rho}\rangle. 
\end{equation}
The trace of the energy-momentum tensor is
\begin{equation}\begin{split}
\langle T_\mu^{\ \mu}\rangle&=\langle-g^{\mu\nu}\partial_\mu\varphi^i\partial_\nu\varphi^i-2m^2(\varphi^i)^2+\frac{2N}{\lambda}\chi^2\rangle\\
&=\langle-\frac{1}{2}\nabla^2(\varphi^i)^2+\varphi^i\nabla^2\varphi^i-2m^2(\varphi^i)^2+\frac{2N}{\lambda}\chi^2\rangle\\
&=-\frac{1}{2}\nabla^2\langle(\varphi^i)^2\rangle-m^2\langle(\varphi^i)^2\rangle+\frac{2N}{\lambda}\chi^2, 
\end{split}\label{4N4}\end{equation}
where $\nabla^2=\frac{1}{\sqrt{-g}}\partial_\mu(\sqrt{-g}g^{\mu\nu}\partial_\nu)$. 
In the third line, we have used the equation of motion
\begin{equation}
\nabla^2\varphi^i-m^2\varphi^i=0. 
\end{equation}
By substituting (\ref{4N1}) and (\ref{4N3}) to (\ref{4N4}), 
\begin{equation}
\langle T_\mu^{\ \mu}\rangle=0,\hspace{1em}\Lambda_{eff}=\Lambda. 
\end{equation}
This result is consistent with that in the stochastic approach (\ref{4sLambda}). 
However this is an exact non-perturbative result in the large $N$ limit beyond the leading logarithmic approximation.
Note that we extract the term which is proportional to $H^2/m^2$ in the right hand side of (\ref{4N2}). 
Subsequent terms give $\mathcal{O}(\lambda^\frac{1}{2})$ shift to the cosmological constant. 

Unlike the scalar field theory with an interaction potential,
we don't know how to evaluate the non-perturbative IR effect in a generic model with derivative interactions. 
Non-linear sigma model is such an example while quantum gravity is another.
It is very important to investigate IR effects in these models.
With this motivation, we consider the non-linear sigma model in the next section. 
We can investigate some non-perturbative effects also since it is exactly solvable in the large $N$ limit. 

\section{Non-linear sigma model}
\setcounter{equation}{0}

As a model with derivative interactions, 
we consider the non-linear sigma model. 
The global symmetry of the model ensures that the non-linear sigma model consists of massless and 
minimally coupled scalar fields. 
Furthermore, it becomes exactly solvable in the large $N$ limit 
which enables us to investigate certain non-perturbative effects.   
For these reasons, we study the non-linear sigma model. 

\subsection{Perturbative result}

First we investigate the non-linear sigma model at the perturbative level. 
The non-linear sigma model in de Sitter space is given by a map from
de Sitter space to a target space with a global continuous symmetry.
The scalar fields $\varphi ^i$  appear as its coordinates representing Nambu-Goldstone bosons
of the spontaneously broken symmetry.
The action is
\begin{equation}
S_{matter}=\frac{1}{2g^2}\int\sqrt{-g}d^4x\ G_{ij}(\varphi)(-g^{\mu\nu}\partial_\mu\varphi^i\partial_\nu\varphi^j), 
\end{equation}
where $g^2$ is the coupling constant 
and $G_{ij}\ (i=1\cdots N)$ is the metric of the target space. 
To investigate the quantum effect at $\mathcal{O}(g^2)$, we use the background field method \cite{Honerkamp1971,AlvarezGaume1981}
\begin{align}
S_{matter}=\frac{1}{2g^2}\int&\sqrt{-g}d^4x\ \Big[G_{ij}(-g^{\mu\nu}\partial_\mu\varphi^i\partial_\nu\varphi^j)
-R_{iajb}(\xi^a\xi^b)(-g^{\mu\nu}\partial_\mu\varphi^i\partial_\nu\varphi^j)\\
&+\big(-g^{\mu\nu}(D_\mu\xi)^a(D_\nu\xi)^a\big)
-\frac{1}{3}R_{cadb}(\xi^a\xi^b)\big(-g^{\mu\nu}(D_\mu\xi)^c(D_\nu\xi)^d\big)+\cdots\Big], \notag
\end{align}
where $\varphi^i$ are the background fields, $\xi^i$ are the quantum fluctuations. 
Here $R_{ikjl}$ is the Riemann tensor and the covariant derivative are
\begin{equation}
D_\mu\xi^i=\partial_\mu\xi^i+\Gamma^i_{\ jk}\partial_\mu\varphi^j\xi^k. 
\end{equation}
By using the vielbein $e_i^{\ a}$, we can work in the flat tangential space $E_{N}$
instead of the target space 
\begin{equation}
\xi^a=e_i^{\ a}\xi^i,\hspace{1em}(D_\mu\xi)^a=\partial_\mu\xi^a+\omega_i^{\ ab}\partial_\mu\varphi^i\xi^b, 
\end{equation}
where $\omega_i^{\ ab}$ is the spin connection. 
Henceforth we rescale the quantum fluctuation $\xi/g\to\xi$ for convenience. 

Here we may assume that the Ricci tensor $R_{ij}$ is proportional to $G_{ij}$ just as
the maximally symmetric space. 
From the terms which are proportional to $G_{ij}(-g^{\mu\nu}\partial_\mu\varphi^i\partial_\nu\varphi^j)$, the effective coupling constant is found as follows
\begin{equation}
\frac{1}{g^2_{eff}}=\frac{1}{g^2}-\frac{R}{N}\langle\xi(x)\xi(x)\rangle-\frac{\delta g^2}{g^4}, 
\end{equation}
where $\langle\xi^a(x)\xi^b(x')\rangle=\delta^{ab}\langle\xi(x)\xi(x')\rangle$, 
$R$ is the Ricci scalar and $\delta g^2$ is the counter term for the coupling constant.  
For example, on the sphere $S_{N}$ whose radius is $1$, 
\begin{equation}
R_{ikjl}=G_{ij}G_{kl}-G_{il}G_{kj},\hspace{1em}
R=(N-1)N>0. 
\end{equation}
The propagator at the coincident point can be read-off from
(\ref{G}), (\ref{alpha}) and (\ref{gamma}), 
\begin{equation}
\langle\xi(x)\xi(x)\rangle
=\frac{1}{4}\left(\frac{H}{2\pi}\right)^24^\frac{\varepsilon}{2}\left(\frac{H}{\sqrt{\pi}}\right)^{-\varepsilon}
\frac{\Gamma(3-\varepsilon)}{\Gamma(2-\frac{\varepsilon}{2})}
\Big\{\delta+2\log a(\tau)\Big\}. 
\end{equation}
In dS space, it contains the time dependent logarithm. 
On the target space where the Ricci tensor is proportional to $G_{ij}$, 
the effective coupling constant is 
\begin{equation}
\frac{1}{g^2_{eff}}=\frac{1}{g^2}-\frac{R}{N}\frac{H^2}{4\pi^2}\log a(\tau), 
\label{pg^2}\end{equation}
\begin{equation}
\delta g^2
=-\frac{R}{N}g^4\times
\frac{1}{4}\left(\frac{H}{2\pi}\right)^24^\frac{\varepsilon}{2}\left(\frac{H}{\sqrt{\pi}}\right)^{-\varepsilon}
\frac{\Gamma(3-\varepsilon)}{\Gamma(2-\frac{\varepsilon}{2})}
\ \delta. 
\label{deltag}\end{equation}
The effective coupling constant increases with the cosmic evolution in the non-linear sigma model on $S_{N}$.
On the other hand, the effective coupling constant decreases with cosmic evolution on a hyperboloid $H_{N}$. 

As is well known, the non-linear sigma model on $S_{N}$ is asymptotically free 
in $2$-dimensional Minkowski space. 
The propagator at the coincident point is  
\begin{equation}
\langle\xi(x)\xi(x)\rangle=\frac{1}{4\pi}\big\{\frac{2}{\varepsilon}-\log\mu^2-\gamma+\log 4\pi\big\}, 
\end{equation}
where $\gamma$ is the Euler's constant. 
We find
\begin{equation}
\frac{1}{g^2_{eff}}=\frac{1}{g^2}+\frac{R}{N}\frac{1}{2\pi}\log \mu. 
\end{equation}
The effective coupling constant increases as the mass scale $\mu$ is decreased in an analogous fashion.
Although there are similarities between the non-linear sigma models in 4 dimensional de Sitter space and 
in 2 dimensional Minkowski space, there are important differences.
Namely the coupling constant in the non-linear sigma model in 4d dS space changes with time while
that in 2d Minkowski space remains the constant. Its evolution takes place under the renormalization
group not under the time evolution.
If the de Sitter invariance is maintained, the time evolution in a comoving coordinate can be related to the
conformal transformation and thus the renormalization group. 
However the de Sitter invariance is broken by the IR quantum effects
\cite{Woodard2008, KK}.

We next consider the contribution to the cosmological constant where we can neglect the background field $\varphi^i$ 
\begin{equation}
\langle T_{\mu\nu}\rangle=
(\delta_\mu^{\ \rho}\delta_\nu^{\ \sigma}-\frac{1}{2}\eta_{\mu\nu}\eta^{\rho\sigma})\langle\partial_\rho\xi^a\partial_\sigma\xi^a
-\frac{g^2}{3}R_{acbd}\partial_\rho\xi^a\partial_\sigma\xi^b\xi^c\xi^d\rangle. 
\label{NLpT}\end{equation}
The one loop contribution is 
\begin{equation}
\langle\partial_\rho\xi^a\partial_\sigma\xi^a\rangle|_{g^0}=-N\frac{3H^4}{32\pi^2}g_{\rho\sigma}. 
\end{equation}
The contributions at $\mathcal{O}(g^2)$ come from the following two terms 
\begin{align}
&-\frac{g^2}{3}R_{acbd}\langle\partial_\rho\xi^a\partial_\sigma\xi^b\xi^c\xi^d\rangle
+(\delta Z-\frac{\delta g^2}{g^2})\langle\partial_\rho\xi^a\partial_\sigma\xi^a\rangle\label{second}\\
=&\ \big\{-\frac{g^2}{3}RG^{++}(x,x)+N(\delta Z-\frac{\delta g^2}{g^2})\big\}
\lim_{x'\to x}\partial_\rho\partial_\sigma' G^{++}(x,x')\notag\\
&+\frac{g^2}{12}R\partial_\rho G^{++}(x,x)\partial_\sigma G^{++}(x,x), \notag
\end{align}
\begin{align}
&\ \langle\partial_\rho\xi^a\partial_\sigma\xi^a\rangle|_{g^2}\label{first}\\
=&\ \int d^Dx'\ a^D(\tau')
\big\{i\frac{g^2}{3}RG^{++}(x',x')-iN(\delta Z-\frac{\delta g^2}{g^2})\big\}\notag\\
&\times g^{\alpha\beta}(\tau')
\big[\partial_\rho\partial_\alpha' G^{++}(x,x')\partial_\sigma\partial_\beta' G^{++}(x,x')
-\partial_\rho\partial_\alpha' G^{+-}(x,x')\partial_\sigma\partial_\beta' G^{+-}(x,x')\big]\notag\\
&+\int d^Dx'\ a^D(\tau')\ 
i\frac{g^2}{3}R\lim_{x''\to x'}\partial_\alpha'\partial_\beta'' G^{++}(x',x'')\notag\\
&\times g^{\alpha\beta}(\tau')
\big[\partial_\rho G^{++}(x,x')\partial_\sigma G^{++}(x,x')
-\partial_\rho G^{+-}(x,x')\partial_\sigma G^{+-}(x,x')\big]\notag\\
&-\int d^Dx'\ a^D(\tau')\ 
i\frac{g^2}{6}R\partial_\alpha' G^{++}(x',x')\notag\\
&\times g^{\alpha\beta}(\tau')\partial_\beta'
\big[\partial_\rho G^{++}(x,x')\partial_\sigma G^{++}(x,x')
-\partial_\rho G^{+-}(x,x')\partial_\sigma G^{+-}(x,x')\big], \notag
\end{align}
where $\delta Z$ is the counter term for the wave function renormalization.
We use the in-in formalism to evaluate these expectation values\cite{Schwinger1961,Keldysh1964}
\begin{equation}
G^{++}(x,x')\equiv\langle T\xi_I(x)\xi_I(x')\rangle,\hspace{1em}
G^{+-}(x,x')\equiv\langle \xi_I(x')\xi_I(x)\rangle. 
\end{equation}
Here $T$ means the time ordering and $\xi_I$ means the interaction picture. 

In each contribution in (\ref{second}) and (\ref{first}), 
there exists  a  propagator which is not affected by the derivatives. 
So there is a single IR log factor in each contribution. 
Note that this power counting argument applies to non-linear sigma models with two derivative interactions. 
At the two loop level, the dominant term in the energy-momentum tensor is proportional to $g_{\mu\nu}g^2\log a(\tau)$. 
Henceforth we calculate only the terms which are proportional to $g_{\mu\nu}g^2\log a(\tau)$ in (\ref{NLpT}). 
To do so, we extract the terms which are proportional to $g_{\rho\sigma}g^2\log a(\tau)$ in (\ref{second}) and 
(\ref{first}). 
As we have explained in the preceding section, there could be no $\delta_\mu^{\ 0}\delta_\nu^{\ 0}g^2\log a(\tau)$ 
type term in the 
energy-momentum tensor due to the conservation law.

We can evaluate (\ref{second}) as 
\begin{equation}\begin{split}
&-\frac{g^2}{3}R_{acbd}\langle\partial_\rho\xi^a\partial_\sigma\xi^b\xi^c\xi^d\rangle
+(\delta Z-\frac{\delta g^2}{g^2})\langle\partial_\rho\xi^a\partial_\sigma\xi^a\rangle\\
\simeq&\ \big\{-\frac{g^2}{3}RG^{++}(x,x)+N(\delta Z-\frac{\delta g^2}{g^2})\big\}
\lim_{x'\to x}\partial_\rho\partial_\sigma' G^{++}(x,x')\\
=&+g_{\rho\sigma}\frac{g^2RH^6}{2^7\pi^4}\log a(\tau). 
\end{split}\label{second1}\end{equation}
If the Ricci tensor is proportional to $G_{ij}$, the wave function renormalization factor is determined 
from (\ref{deltag}) as
\begin{equation}
\delta Z=
-\frac{R}{N}g^2\times
\frac{1}{6}\left(\frac{H}{2\pi}\right)^24^\frac{\varepsilon}{2}\left(\frac{H}{\sqrt{\pi}}\right)^{-\varepsilon}
\frac{\Gamma(3-\varepsilon)}{\Gamma(2-\frac{\varepsilon}{2})}
\ \delta. 
\end{equation}

(\ref{first}) is estimated as
\begin{equation}\begin{split}
&\ \langle\partial_\rho\xi^a\partial_\sigma\xi^a\rangle|_{g^2}\\
\simeq&\ \int d^Dx'\ a^D(\tau')
\big\{i\frac{g^2}{3}RG^{++}(x',x')-iN(\delta Z-\frac{\delta g^2}{g^2})\big\}\\
&\times g^{\alpha\beta}(\tau')
\big[\partial_\rho\partial_\alpha' G^{++}(x,x')\partial_\sigma\partial_\beta' G^{++}(x,x')
-\partial_\rho\partial_\alpha' G^{+-}(x,x')\partial_\sigma\partial_\beta' G^{+-}(x,x')\big]\\
=&\ i\frac{2g^2}{3}R\alpha\beta\int d^{4-\varepsilon}x'\ a^{4-\varepsilon}(\tau')\log a(\tau')\\
&\times g^{\alpha\beta}(\tau')
\big[\partial_\rho\partial_\alpha' G^{++}(x,x')\partial_\sigma\partial_\beta' G^{++}(x,x')
-\partial_\rho\partial_\alpha' G^{+-}(x,x')\partial_\sigma\partial_\beta' G^{+-}(x,x')\big]. 
\end{split}\label{first1}\end{equation}
Here the integrand is written as follows using the parametrization of the propagator given in (\ref{G}) 
\begin{equation}\begin{split}
&g^{\alpha\beta}(\tau')\partial_\rho\partial_\alpha' G(x,x')\partial_\sigma\partial_\beta' G(x,x')\\
=&\alpha^2a^{-2}(\tau')\eta^{\alpha\beta}\big\{\gamma'^2(y)\partial_\rho \partial_\alpha' y\partial_\sigma \partial_\beta' y
+\gamma''^2(y)\partial_\rho y\partial_\alpha' y\partial_\sigma y\partial_\beta' y\\
&\hspace{5em}+\gamma''(y)\gamma'(y)(\partial_\rho \partial_\alpha' y\partial_\sigma y\partial_\beta' y+\partial_\rho y\partial_\alpha' y\partial_\sigma \partial_\beta' y)\big\}.
\end{split}\end{equation}

By substituting explicit expressions in (\ref{alpha}) and (\ref{gamma}) to (\ref{first1})  and
after neglecting the terms which are proportional to $\delta_\rho^{\ 0}$ or $\delta_\sigma^{\ 0}$, we obtain
\begin{equation}\begin{split}
&\ \langle\partial_\rho\xi^a\partial_\sigma\xi^a\rangle|_{g^2}\\
\simeq&\ i\frac{2g^2}{3}R\alpha^3\beta H^4a^2(\tau)\int d^{4-\varepsilon}x'a^{4-\varepsilon}(\tau')\log a(\tau')\sum_m F^m_{\rho\sigma}\\
=&\ i\frac{g^2}{3}R\left(\frac{H}{2\pi}\right)^6a^2(\tau)
\times H^4A\int d^{4-\varepsilon}x'a^{4-\varepsilon}(\tau')\log a(\tau')\sum_m F^m_{\rho\sigma},
\end{split}\end{equation}
\begin{equation}
A\equiv 4^\frac{\varepsilon}{2}\left(\frac{H}{\sqrt{\pi}}\right)^{-3\varepsilon}(1-\frac{\varepsilon}{2})^2(1-\varepsilon)\Gamma(1-\frac{\varepsilon}{2})\Gamma(1-\varepsilon).
\end{equation}
Here we have introduce the following eight tensors:
\begin{equation}
F^1_{\rho\sigma}\equiv\eta_{\rho\sigma}\big[\frac{4}{y^{4-\varepsilon}_{++}}
-\frac{4}{y^{4-\varepsilon}_{+-}}\big], 
\end{equation}
\begin{equation}
F^2_{\rho\sigma}\equiv\eta_{\rho\sigma}\big[\frac{4(1-\frac{\varepsilon}{4})}{y^{3-\varepsilon}_{++}}
-\frac{4(1-\frac{\varepsilon}{4})}{y^{3-\varepsilon}_{+-}}\big], 
\end{equation}
\begin{equation}
F^3_{\rho\sigma}\equiv\eta_{\rho\sigma}\big[\frac{(1-\frac{\varepsilon}{4})^2}{y^{2-\varepsilon}_{++}}
-\frac{(1-\frac{\varepsilon}{4})^2}{y^{2-\varepsilon}_{+-}}\big],
\end{equation}
\begin{equation}
F^4_{\rho\sigma}\equiv
-\frac{1}{2}\eta_{\rho\sigma}
\left(\frac{\Gamma(4-\varepsilon)}{\Gamma(3-\frac{\varepsilon}{2})\Gamma(2-\frac{\varepsilon}{2})}
\big[\frac{4^\frac{\varepsilon}{2}}{y^{2-\frac{\varepsilon}{2}}_{++}}
-\frac{4^\frac{\varepsilon}{2}}{y^{2-\frac{\varepsilon}{2}}_{+-}}\big]
-\frac{\Gamma(4-\frac{\varepsilon}{2})}{2\Gamma(2-\frac{\varepsilon}{2})}
\big[\frac{1}{y^{2-\varepsilon}_{++}}-\frac{1}{y^{2-\varepsilon}_{+-}}\big]\right), 
\end{equation}
\begin{equation}
F^5_{\rho\sigma}\equiv
\Delta x_\rho\Delta x_\sigma\cdot 32(1-\frac{3}{4}\varepsilon)H^2a^2(\tau')
\big[\frac{1}{y^{5-\varepsilon}_{++}}-\frac{1}{y^{5-\varepsilon}_{+-}}\big], 
\end{equation}
\begin{equation}
F^6_{\rho\sigma}\equiv
\Delta x_\rho\Delta x_\sigma\cdot 4(1-\frac{7}{2}\varepsilon)H^2a^2(\tau')
\big[\frac{1}{y^{4-\varepsilon}_{++}}-\frac{1}{y^{4-\varepsilon}_{+-}}\big], 
\end{equation}
\begin{equation}
F^7_{\rho\sigma}\equiv
\Delta x_\rho\Delta x_\sigma\cdot -\frac{9}{2}\varepsilon H^2a^2(\tau')
\big[\frac{1}{y^{3-\varepsilon}_{++}}-\frac{1}{y^{3-\varepsilon}_{+-}}\big], 
\end{equation}
\begin{equation}\begin{split}
F^8_{\rho\sigma}\equiv&
\Delta x_\rho\Delta x_\sigma\cdot H^2a^2(\tau')\frac{4-\varepsilon}{2}\\
&\times\left(\frac{\Gamma(4-\varepsilon)}{\Gamma(3-\frac{\varepsilon}{2})\Gamma(2-\frac{\varepsilon}{2})}
\big[\frac{4^\frac{\varepsilon}{2}}{y^{3-\frac{\varepsilon}{2}}_{++}}
-\frac{4^\frac{\varepsilon}{2}}{y^{3-\frac{\varepsilon}{2}}_{+-}}\big]
-\frac{\Gamma(4-\frac{\varepsilon}{2})}{2\Gamma(2-\frac{\varepsilon}{2})}
\big[\frac{1}{y^{3-\varepsilon}_{++}}-\frac{1}{y^{3-\varepsilon}_{+-}}\big]\right), 
\end{split}\end{equation}
where $\Delta x^\rho$, $y_{++}$ and $y_{+-}$ are
\begin{equation}\begin{split}
&\Delta x_\rho=x_\rho-x'_\rho, \\
&\Delta x^2_{++}=-(|\tau-\tau'|-ie)^2+({\bf x}-{\bf x}')^2, \\
&\Delta x^2_{+-}=-(\tau-\tau'+ie)^2+({\bf x}-{\bf x}')^2,
\end{split}\label{in-in}\end{equation}
\begin{equation}\begin{split}
y_{++}=H^2a(\tau)a(\tau')\Delta x^2_{++},\hspace{1em}y_{+-}=H^2a(\tau)a(\tau')\Delta x^2_{+-}. 
\end{split}\end{equation}
We explain how to calculate these integrals containing $F_{\rho\sigma}^m$ in Appendix A. 

We simply lists the results:
\begin{equation}\begin{split}
&\ H^4A\int d^{4-\varepsilon}x'a^{4-\varepsilon}(\tau')\log a(\tau')F^1_{\rho\sigma}\\
\simeq&\ 4i\pi^2\log a(\tau)\ \eta_{\rho\sigma}\cdot 0, 
\end{split}\end{equation}
\begin{equation}\begin{split}
&\ H^4A\int d^{4-\varepsilon}x'a^{4-\varepsilon}(\tau')\log a(\tau')F^2_{\rho\sigma}\\
\simeq&\ 4i\pi^2\log a(\tau)\ \eta_{\rho\sigma}\cdot\left\{\frac{1}{2}\frac{\zeta}{\varepsilon}+\frac{1}{2}\log\frac{2\mu}{H}+\frac{1}{8}\right\}, 
\end{split}\end{equation}
\begin{equation}\begin{split}
&\ H^4A\int d^{4-\varepsilon}x'a^{4-\varepsilon}(\tau')\log a(\tau')F^3_{\rho\sigma}\\
\simeq&\ 4i\pi^2\log a(\tau)\ \eta_{\rho\sigma}\cdot\left\{-\frac{1}{2}\frac{\zeta}{\varepsilon}-\frac{1}{2}\log\frac{2\mu}{H}+\frac{1}{4}\right\}, 
\end{split}\end{equation}
\begin{equation}\begin{split}
&\ H^4A\int d^{4-\varepsilon}x'a^{4-\varepsilon}(\tau')\log a(\tau')F^5_{\rho\sigma}\\
\simeq&\ 4i\pi^2\log a(\tau)\ \eta_{\rho\sigma}\cdot\left\{-\frac{1}{4}\frac{\zeta}{\varepsilon}-\frac{1}{4}\log\frac{2\mu}{H}+0\right\}, 
\end{split}\end{equation}
\begin{equation}\begin{split}
&\ H^4A\int d^{4-\varepsilon}x'a^{4-\varepsilon}(\tau')\log a(\tau')F^6_{\rho\sigma}\\
\simeq&\ 4i\pi^2\log a(\tau)\ \eta_{\rho\sigma}\cdot\left\{\frac{1}{4}\frac{\zeta}{\varepsilon}+\frac{1}{4}\log\frac{2\mu}{H}-\frac{3}{4}\right\}, 
\end{split}\end{equation}
\begin{equation}\begin{split}
&\ H^4A\int d^{4-\varepsilon}x'a^{4-\varepsilon}(\tau')\log a(\tau')F^7_{\rho\sigma}\\
\simeq&\ 4i\pi^2\log a(\tau)\ \eta_{\rho\sigma}\cdot\frac{9}{16}, 
\end{split}\end{equation}
\begin{equation}\begin{split}
&\ H^4A\int d^{4-\varepsilon}x'a^{4-\varepsilon}(\tau')\log a(\tau')(F^4_{\rho\sigma}+F^8_{\rho\sigma})\\
\simeq&\ 4i\pi^2\log a(\tau)\ \eta_{\rho\sigma}\cdot\frac{3}{16}, 
\end{split}\end{equation}
where $\zeta$ is
\begin{equation}
\zeta\equiv\left(\frac{2\pi}{\mu H}\right)^\varepsilon(1-\frac{\varepsilon}{2})^2(1-\varepsilon)\Gamma(1-\varepsilon). 
\end{equation}
The total of these eight contributions is
\begin{equation}
H^4A\int d^{4-\varepsilon}x'a^{4-\varepsilon}(\tau')\log a(\tau')\sum_mF^m_{\rho\sigma}
\simeq 4i\pi^2\log a(\tau)\ \eta_{\rho\sigma}\cdot\frac{3}{8}. 
\end{equation}
In this way, the quantum expectation value of the quadratic kinetic term is found as follows
up to the 2 loop level
\begin{equation}
\langle\partial_\rho\xi^a\partial_\sigma\xi^a\rangle\simeq 
-g_{\rho\sigma}N\frac{3H^4}{32\pi^2}-g_{\rho\sigma}\frac{g^2RH^6}{2^7\pi^4}\log a(\tau). 
\label{first2}\end{equation} 
Note that unlike in the scalar field theory with an interaction potential, 
the expectation value of the quadratic kinetic term $\langle\partial_\rho\xi^a\partial_\sigma\xi^a\rangle$
is of the same order with the quartic interaction term 
$-\frac{g^2}{3}R_{acbd}\langle\partial_\rho\xi^a\partial_\sigma\xi^b\xi^c\xi^d\rangle$.

By combining (\ref{second1}), (\ref{first2}), we find that there is no time dependence of the vev of the energy-momentum tensor up to the two loop level.
\begin{equation}
\langle T_{\mu\nu}\rangle\simeq N\frac{3H^4}{32\pi^2}g_{\mu\nu}. 
\end{equation}
Although there are time dependent IR logarithms in each contribution in agreement with the
power counting arguments, they cancel out each other . 
The contribution to the cosmological constant is identical to that in the free field theory.  
\begin{equation}
\Lambda_{eff}\simeq\Lambda-\kappa N\frac{3H^4}{32\pi^2}. 
\label{NLpLambda}\end{equation}
Note that we have neglected the term which is proportional to $g_{\mu\nu}g^2$. 
Such a term has no time dependence and can be subtracted by a counter term. 

\subsection{Non-perturbative result : In the large $N$ limit}

The IR power counting suggests that there could arise a $(\log a )^{n-1}$ factor at the $n$-th 
loop level in the expectation value of the energy-momentum tensor.
It is because there could be $n-1$ propagators left intact by the derivatives in a relevant diagram. 
At late times they could lead to large non-perturbative effects.
In order to understand them,
we investigate the non-linear sigma model on a sphere $S_{N}$ in the large $N$ limit. 
The action of the non-linear sigma model on $S_{N}$ is written as follows by introducing an auxiliary field $\chi$
\begin{equation}
S_{matter}=\int\sqrt{-g}d^4x\ \big[-\frac{1}{2}g^{\mu\nu}\partial_\mu\varphi^i\partial_\nu\varphi^i
-\frac{\chi}{2}\big((\varphi^i)^2-\frac{1}{g^2}\big)\big], 
\label{NLN}\end{equation}
where $i=1\cdots N+1$. 
The field $\chi$ imposes the following constraint 
\begin{equation}
(\varphi^i)^2=\frac{1}{g^2}. 
\label{NLN1}\end{equation}

In the large $N$ limit, we can neglect the fluctuation of $\chi$. 
In the early stage, the IR quantum fluctuation of a scalar field is small.
In order to satisfy the constraint (\ref{NLN1}), 
we introduce the classical expectation value $(\varphi_{cl}^i(x))^2$ : 
\begin{equation}
(\varphi_{cl}^i(x))^2+\langle(\varphi^i(x))^2\rangle=\frac{1}{g^2}. 
\label{NLN2}\end{equation}
It is because $1/g^2$ is a constant 
and the propagator for scalar fields is time dependent 
\begin{equation}
\langle(\varphi^i(x))^2\rangle= (N+1)\frac{H^2} {4\pi^2}\log a(\tau). 
\label{NLN3}\end{equation}
The classical expectation value $(\varphi_{cl}^i(x))^2$ is identified with the effective coupling constant : 
\begin{equation}
(\varphi_{cl}^i(x))^2\equiv\frac{1}{g^2_{eff}}. 
\label{NLN4}\end{equation}
From (\ref{NLN2}), (\ref{NLN3}) and (\ref{NLN4}), 
\begin{equation}
\frac{1}{g^2_{eff}}=\frac{1}{g^2}-(N+1)\frac{H^2} {4\pi^2}\log a(\tau). 
\label{Ng^2}\end{equation}
The effective coupling constant increases with time. 
It agrees with the one loop result (\ref{pg^2}) to the leading order in $N$. 
Since $\varphi_{cl}^i(x)$ is a free field with $m^2=\chi$, we can perturbatively determine its mass from (\ref{NLN4}) 
and the equations of motion as 
\begin{equation}\begin{split}
m^2=\chi&=(\varphi_{cl}^i)^{-1}\nabla^2\varphi^i_{cl}\\
&\simeq\frac{3(N+1)g^2H^4}{8\pi^2}. 
\end{split}\label{NLN6}\end{equation}
Here we don't perform the summation over $i$. 

As the dS expansion continues, the effective coupling becomes strong.
In such a limit, the IR singularity is avoided by the dynamical mass generation we have demonstrated. 
In this case, the action (\ref{NLN}) is identified as the free massive field theory with the constant term $\chi/2g^2$. 
The propagator of the massive fields at the coincident point is
\begin{equation}
\langle(\varphi^i)^2\rangle=(N+1)\frac{3H^4}{8\pi^2m^2}, 
\label{NLN5}\end{equation}
where we have assumes again that $m^2/H^2$ is small.
From (\ref{NLN2}) and (\ref{NLN5}), the mass of the scalar field is determined precisely
as (\ref{NLN6}).
Such a dynamical mass generation mechanism is pointed out in \cite{Davis1991}. 
We have demonstrated that the effective coupling constant increases as time goes on 
and the dynamical mass generation indeed takes place. 
The trace of the energy-momentum tensor is evaluated as follows 
\begin{equation}\begin{split}
\langle T_\mu^{\ \mu}\rangle&=\langle-g^{\mu\nu}\partial_\mu\varphi^i\partial_\nu\varphi^i-2m^2(\varphi^i)^2+\frac{2\chi}{g^2}\rangle\\
&=\langle-\frac{1}{2}\nabla^2(\varphi^i)^2+\varphi^i\nabla^2\varphi^i-2m^2(\varphi^i)^2+\frac{2\chi}{g^2}\rangle\\
&=-\frac{1}{2}\nabla^2\langle(\varphi^i)^2\rangle-m^2\langle(\varphi^i)^2\rangle+\frac{2\chi}{g^2}. 
\label{NLN7}\end{split}\end{equation}
In the third line, we have used the equation of motion 
\begin{equation}
\nabla^2\varphi^i-m^2\varphi^i=0. 
\end{equation}
By substituting (\ref{NLN5}) and (\ref{NLN6}) to (\ref{NLN7}), we obtain
\begin{equation}
\langle T_\mu^{\ \mu}\rangle=(N+1)\frac{3H^4}{8\pi^2},\hspace{1em}\Lambda_{eff}=\Lambda-\kappa (N+1)\frac{3H^4}{32\pi^2}. 
\label{NLNLambda2}\end{equation}
The effective cosmological constant is identical to the perturbative one (\ref{NLpLambda}) to the leading order in $N$.
We conclude that the interaction in the non-linear sigma model does not contribute
to the cosmological constant in the large $N$ limit. 
Of course, there are subsequent $\mathcal{O}(g^2N)$ terms 
but they are negligible if $g^2N\ll H^{-2}$. 
We have found a complete cancellation of the IR logarithms which is not expected from the IR power counting arguments.

We summarize the results of our investigation on the non-linear sigma model in 4d dS space. 
As it is shown in (\ref{NLpLambda}), 
the interaction doesn't induce time dependence to the cosmological constant at the 2-loop level. 
In the large $N$ limit, 
the effective cosmological constant is identical to that of the free fields. 
Even after the dynamical mass generation takes place, 
the contribution to the cosmological constant remains the same with the free massless field theory. 
Contrary to the IR power counting arguments, we find that the interaction doesn't make the cosmological constant 
time dependent in the non-linear sigma model in an explicit perturbation theory and large $N$ limit.

In $\varphi^4$ theory, the interaction contributes to the cosmological constant mainly by the potential term. 
The expectation value of the potential term increases as time goes on and finally approaches a constant value. 
On the other hand, the non-linear sigma model is found to be equivalent to the free massless field theory 
with respect to the contribution to the cosmological constant.

\section{Trace of the energy-momentum tensor}

In this section, we evaluate the effective cosmological constant by using the equation of motion. 
We can check the validity of our results in the preceding sections by simply evaluating the trace part
of the energy-momentum tensor. 
 It is because the
 $g_{\mu\nu}$ term is always dominant in the energy-momentum tensor 
irrespectively whether the de Sitter invariance is respected or
 broken by IR quantum effects.
So the vev of the energy-momentum tensor is written as follows to the leading order of the IR effect 
\begin{equation}
\langle T_{\mu\nu}\rangle\simeq \frac{g_{\mu\nu}}{4}\langle T_\rho^{\ \rho}\rangle. 
\end{equation}
The effective cosmological constant is 
\begin{equation}
\Lambda_{eff}\simeq\Lambda-\frac{\kappa}{4}\langle T_\mu^{\ \mu}\rangle. 
\end{equation}

Let us consider $\varphi^4$ theory, 
\begin{equation}\begin{split}
\langle T_\mu^{\ \mu}\rangle
&=\langle-g^{\mu\nu}\partial_\mu\varphi\partial_\nu\varphi-\frac{\lambda}{6}\varphi^4\rangle\\
&=\langle-\frac{1}{2}\nabla^2(\varphi^2)+\varphi\nabla^2\varphi-\frac{\lambda}{6}\varphi^4\rangle\\
&=-\frac{1}{2}\nabla^2\langle\varphi^2\rangle. 
\label{4T}\end{split}\end{equation}
In the third line, we have used the equation of motion 
\begin{equation}
\nabla^2\varphi-\frac{\lambda}{6}\varphi^3=0.  
\end{equation}
Perturbatively, $\langle\varphi^2\rangle$ is evaluated as follows \cite{Woodard2005} up to the 2 loop level
\begin{equation}
\langle\varphi^2\rangle\simeq\frac{H^2}{4\pi^2}\log a(\tau)-\frac{\lambda H^2}{2^4\cdot 3^2\pi^4}\log^3 a(\tau). 
\label{quadratic3}\end{equation}
By substituting (\ref{quadratic3}) to (\ref{4T}), 
\begin{equation}\begin{split}
\langle T_\mu^{\ \mu}\rangle&\simeq\frac{3H^4}{8\pi^2}-\frac{\lambda H^4}{2^5\pi^4}\log^2 a(\tau),\\
\Lambda_{eff}&\simeq\Lambda-\kappa\frac{3H^4}{32\pi^4}+\kappa \frac{\lambda H^4}{2^7\pi^4}\log^2a(\tau). 
\end{split}\end{equation}
This result agrees with (\ref{4pLambda}). 

In the stochastic approach, $\langle\varphi^2\rangle$ eventually becomes constant. 
Therefore, from (\ref{4T}), 
\begin{equation}
\langle T_\mu^{\ \mu}\rangle=0,\hspace{1em}
\Lambda_{eff}=\Lambda. 
\end{equation}
The contribution from a scalar field to the effective cosmological constant vanishes. 
This result is in agreement with (\ref{4sLambda}). 
It is an accident in $\varphi^4$ theory. 
In $\varphi^{2m}\ (m\not=2)$ theory, the contribution from the potential term remains after using the equation of motion. 

The non-linear sigma model can be investigated in a similar way. 
In a perturbative expansion, the trace of the energy-momentum tensor is evaluated as
\begin{equation}\begin{split}
\langle T_\mu^{\ \mu}\rangle
=&\ \langle-g^{\mu\nu}\partial_\mu\xi^a\partial_\nu\xi^a
+\frac{g^2}{3}R_{acbd}g^{\mu\nu}\partial_\mu\xi^a\partial_\nu\xi^b\xi^c\xi^d \rangle\\
=&\ \langle-\frac{1}{2}\nabla^2(\xi^a\xi^a)+\xi^a\nabla^2\xi^a
+\frac{g^2}{3}R_{acbd}g^{\mu\nu}\partial_\mu\xi^a\partial_\nu\xi^b\xi^c\xi^d \rangle\\
=&\ \langle-\frac{1}{2}\nabla^2(\xi^a\xi^a)
+\frac{g^2}{6}(R_{acbd}+R_{bcad})\xi^a\frac{1}{\sqrt{-g}}\partial_\mu(\sqrt{-g}g^{\mu\nu}\partial_\nu\xi^b\xi^c\xi^d) \rangle. 
\end{split}\end{equation}
In the third line, we have used the equation of motion
\begin{equation}\begin{split}
\nabla^2\xi^a
-\frac{g^2}{6}(R_{acbd}+R_{bcad})\frac{1}{\sqrt{-g}}\partial_\mu(\sqrt{-g}g^{\mu\nu}\partial_\nu\xi^b\xi^c\xi^d)&\\
+\frac{g^2}{6}(R_{cadb}+R_{cbda})g^{\mu\nu}\partial_\mu\xi^c\partial_\nu\xi^d\xi^b&=0. 
\end{split}\end{equation}
We then use the equation of motion at $\mathcal{O}(g^0)$ 
and extract the dominant terms with respect to the IR logarithms, that is, the term which contains $\langle\xi\xi\rangle$
\begin{equation}\begin{split}
\langle T_\mu^{\ \mu}\rangle
\simeq&-\frac{1}{2}\nabla^2\langle\xi^a\xi^a\rangle
+\frac{g^2}{3}R_{acbd}\langle\xi^a\xi^d\rangle \langle g^{\mu\nu}\partial_\nu\xi^b\partial_\mu\xi^c \rangle\\
=&-\frac{1}{2}\nabla^2\langle\xi^a\xi^a\rangle+\frac{g^2RH^6}{2^5\pi^4}\log a(\tau). 
\end{split}\label{NLtrace}\end{equation}

The propagator at the coincident point is evaluated as follows to $\mathcal{O}(g^2)$ 
\begin{equation}
\langle\xi^a\xi^a\rangle\simeq N\frac{H^2}{4\pi^2}\log a(\tau)-\frac{g^2RH^4}{2^5\cdot 3\pi^4}\log^2 a(\tau). 
\label{NLpquadratic}\end{equation}
We explain how to calculate it in Appendix B. 
From (\ref{NLtrace}) and (\ref{NLpquadratic}), 
we find that the contribution to the cosmological constant has no time dependence up to the two loop level. 
\begin{equation}\begin{split}
\langle T_\mu^{\ \mu}\rangle&\simeq N\frac{3H^4}{8\pi^2},\\
\Lambda_{eff}&\simeq\Lambda-\kappa N\frac{3H^4}{32\pi^2}. 
\end{split}\end{equation}
This result is in agreement with (\ref{NLpLambda}). 

\section{Conclusion}
\setcounter{equation}{0}

In a quantum field theory in dS space containing a massless and minimally coupled scalar field, we need to introduce 
an IR cut-off to regularize the IR divergence in the propagator.
In other words, there is no de Sitter invariant vacuum in such a theory.
This IR cut-off breaks the dS invariance and gives rise to a growing time dependence to the propagator. 
In an interacting field theory, it could induce a time dependence 
to the effective cosmological constant. We have investigated such an effect
by evaluating the quantum expectation value of the energy-momentum tensor. 

In a scalar field theory interacting through the potential, 
the potential term dominantly contributes to the 
effective cosmological constant in comparison to the kinetic term. 
Since the IR logarithms become large with the cosmic expansion, 
we need to resum these logarithms to all orders.
The leading IR effect, including the non-perturbative effect, can be evaluated by the stochastic approach. 
Although the expectation value of the energy-momentum tensor does not respect de Sitter invariance,
the two independent coefficients are related by the conservation law. 

In this paper, as a model with derivative interactions, 
we have investigated the non-linear sigma model in perturbation theory and in the large $N$ limit. 
In agreement with the IR power counting,  the coupling constant of the non-linear sigma model becomes time dependent at the one loop level.
The same argument indicates that there could arise a $(\log a)^{n-1}$ factor at the $n$-th loop level in the expectation value of the energy momentum tensor.
In perturbation theory, we have indeed found that such an IR log factor in each contribution at the two loop
level. 

However the IR logs at the two loop level
cancel each other to give no time dependence to the effective cosmological constant.
The time independence of the effective cosmological constant is also found at a non-perturbative level.
We indeed find that the cosmological constant is not renormalized by the interaction in the large $N$ limit.
We may reflect on our results in this paper as follows.
If the non-linear sigma model approaches an equilibrium state at late times, it
may be described by a Euclidean field theory on $S_4$.
IR dynamics of such a theory is trivial since the zero mode decouples from the action as it represents
a global symmetry transformation.

Although large $N$ limit is available in the non-linear sigma model,
we don't know how to evaluate the non-perturbative IR effect in
a general model with derivative interactions, 
It is very desirable to develop such a tool. 
Especially such a tool is necessary to understand the IR quantum effects of gravity. 
That is because the gravitational field contains massless and minimally coupled modes
\cite{Woodard1996,Tanaka2007,Tsamis2007,Urakawa2010}. 
The existence of an equilibrium state is also questionable as
the Euclidean quantum gravity suffers from the conformal mode instability.


\section*{Acknowledgments}

\hspace{0.7cm}
This work is supported in part by Grant-in-Aid for Scientific Research from
the Ministry of Education, Science and Culture of Japan.
We would like to thank A. Ishibashi, S. Iso,  H. Kodama, D. Lyth,  J. Nishimura, A. Starobinsky and T. Tanaka for discussions
and information.

\appendix

\section{Integrals containing $F_{\rho\sigma}^m$ tensors}
\setcounter{equation}{0}
Here we explain how to calculate the integrals containing $F_{\rho\sigma}^m$ tensors. 
Most of them can be evaluated by applying the procedure developed in  \cite{Woodard2002}.
However special considerations are required for the tensors with $m=4, 8$. 

\subsection{Integrals containing $F_{\rho\sigma}^m$ except with $m=4, 8$}

First, we calculate the integral which contain $F_{\rho\sigma}^1$. 
The integral of $F_{\rho\sigma}^1$ is
\begin{equation}\begin{split}
&\ H^4A\int d^{4-\varepsilon}x'\ a^{4-\varepsilon}(\tau')\log a(\tau')F^1_{\rho\sigma}\\
=&\ \eta_{\rho\sigma}H^{-4+2\varepsilon}Aa^{-4+\varepsilon}(\tau)
\int d^{4-\varepsilon}x'\ \log a(\tau')\times 4\big[\frac{1}{\Delta x^{8-2\varepsilon}_{++}}-\frac{1}{\Delta x^{8-2\varepsilon}_{+-}}\big]. 
\end{split}\label{A1}\end{equation}
The integrand is written as follows
\begin{equation}
\frac{1}{\Delta x^{8-2\varepsilon}}=\frac{-1}{2^5(3-\varepsilon)(2-\varepsilon)(1-\varepsilon)(2-\frac{\varepsilon}{2})(1-\frac{\varepsilon}{2})}
\partial^4\frac{\partial^2}{\varepsilon}\frac{1}{\Delta x^{2-2\varepsilon}}, 
\label{A2}\end{equation}
where we abbreviate the indexes $++,+-$ because the above identities work out in both cases. 
From (\ref{in-in}), 
\begin{equation}\begin{split}
&\partial^2\frac{1}{\Delta x^{2-\varepsilon}_{++}}=\frac{2ie(2-\varepsilon)\delta(\tau-\tau')}{\big(({\bf x}-{\bf x}')^2+e^2\big)^{2-\frac{\varepsilon}{2}}}
\to\frac{4i\pi^{2-\frac{\varepsilon}{2}}}{\Gamma(1-\frac{\varepsilon}{2})}\delta^{(D)}(x-x'), \\
&\partial^2\frac{1}{\Delta x^{2-\varepsilon}_{+-}}=0. 
\end{split}\label{A3}\end{equation}
By using (\ref{A3}), we extract the UV divergent part 
\begin{equation}\begin{split}
\frac{\partial^2}{\varepsilon}\frac{1}{\Delta x^{2-2\varepsilon}_{++}}
=&\frac{\partial^2}{\varepsilon}\big\{\frac{1}{\Delta x^{2-2\varepsilon}_{++}}-\frac{\mu^{-\varepsilon}}{\Delta x^{2-\varepsilon}_{++}}\big\}
+\frac{4i\pi^{2-\frac{\varepsilon}{2}}\mu^{-\varepsilon}}{\varepsilon\Gamma(1-\frac{\varepsilon}{2})}\delta^{(D)}(x-x')\\
=&\frac{\partial^2}{2}\big\{\frac{\log(\mu^2\Delta x^2_{++})}{\Delta x^2_{++}}\big\}
+\frac{4i\pi^{2-\frac{\varepsilon}{2}}\mu^{-\varepsilon}}{\varepsilon\Gamma(1-\frac{\varepsilon}{2})}\delta^{(D)}(x-x'), \\
\frac{\partial^2}{\varepsilon}\frac{1}{\Delta x^{2-2\varepsilon}_{+-}}
=&\frac{\partial^2}{\varepsilon}\big\{\frac{1}{\Delta x^{2-2\varepsilon}_{+-}}-\frac{\mu^{-\varepsilon}}{\Delta x^{2-\varepsilon}_{+-}}\big\}\\
=&\frac{\partial^2}{2}\big\{\frac{\log(\mu^2\Delta x^2_{+-})}{\Delta x^2_{+-}}\big\}, 
\end{split}\label{A4}\end{equation}
where we introduce the mass parameter $\mu$ to correct the dimension. 

By substituting (\ref{A2}) and (\ref{A4}) to (\ref{A1}), 
\begin{equation}\begin{split}
&\ H^4A\int d^{4-\varepsilon}x'\ a^{4-\varepsilon}(\tau')\log a(\tau')F^1_{\rho\sigma}\\
=&\ \eta_{\rho\sigma}\frac{-1}{96}(1+\frac{31}{12}\varepsilon)H^{-4+2\varepsilon}Aa^{-4+\varepsilon}(\tau)\partial_0^4
\int d^{4-\varepsilon}x'\ \log a(\tau')\\
&\times\left\{\frac{\partial^2}{2}\big[\frac{\log(\mu^2\Delta x^2_{++})}{\Delta x^2_{++}}
-\frac{\log(\mu^2\Delta x^2_{+-})}{\Delta x^2_{+-}}\big]
+\frac{4i\pi^{2-\frac{\varepsilon}{2}}\mu^{-\varepsilon}}{\varepsilon\Gamma(1-\frac{\varepsilon}{2})}\delta^{(D)}(x-x')\right\}, 
\end{split}\label{A5}\end{equation}
where the derivative operator outside the integral is equal to the time derivative $\partial_\alpha\to\delta_\alpha^{\ 0}\partial_0$. 
To evaluate (\ref{A5}), we use the following identity
\begin{equation}
\frac{\log(\mu^2\Delta x^2)}{\Delta x^2}=\frac{1}{8}\partial^2\big\{\log^2(\mu^2\Delta x^2)-2\log(\mu^2\Delta x^2)\big\}. 
\label{A6}\end{equation}
From (\ref{in-in}), the each logarithm is 
\begin{equation}\begin{split}
\log(\mu^2\Delta x^2_{++})&=\log(\mu^2|\Delta \tau^2-r^2|)+i\pi\theta(\Delta \tau^2-r^2), \\
\log(\mu^2\Delta x^2_{+-})&=\log(\mu^2|\Delta \tau^2-r^2|)-i\pi\theta(\Delta \tau^2-r^2)\{\theta(\Delta\tau)-\theta(-\Delta\tau)\}, 
\end{split}\label{A7}\end{equation}
where $\Delta \tau=\tau-\tau',\ r\equiv|{\bf x}-{\bf x}'|$. 
By using (\ref{A6}) and (\ref{A7}), 
\begin{align}
&\int d^4x'\ \log a(\tau')\frac{\partial^2}{2}
\big[\frac{\log(\mu^2\Delta x^2_{++})}{\Delta x^2_{++}}-\frac{\log(\mu^2\Delta x^2_{+-})}{\Delta x^2_{+-}}\big]\label{A8}\\
=&\ i\pi^2\partial_0^4\int^\tau_{-\frac{1}{H}}d\tau'\ \log a(\tau')\int^{\Delta \tau}_0r^2dr\ \big\{\log\big(\mu^2(\Delta\tau^2-r^2)\big)-1\big\}\notag\\
=&\ i\pi^2\partial_0^4\int^\tau_{-\frac{1}{H}}d\tau'\ \log a(\tau')\Delta\tau^3\big\{\frac{2}{3}\log(2\mu\Delta\tau)-\frac{11}{9}\big\}\notag\\
=&\ 4i\pi^2\partial_0\int^\tau_{-\frac{1}{H}}d\tau'\ \log a(\tau')\log(2\mu\Delta\tau)\notag\\
=&\ 4i\pi^2\Big\{\log a(\tau)\log\frac{2\mu}{H}-\log^2a(\tau)
-a^2(\tau)\frac{\partial}{\partial a(\tau)}\int^{a(\tau)}_1da(\tau')\sum^\infty_{n=1}\frac{a^{n-2}(\tau')}{na^n(\tau)}\log a(\tau')\Big\}. \notag
\end{align}
By substituting (\ref{A8}) to (\ref{A5}), 
\begin{align}
&\ H^4A\int d^{4-\varepsilon}x'\ a^{4-\varepsilon}(\tau')\log a(\tau')F^1_{\rho\sigma}\label{A9}\\
=&\ 4i\pi^2\eta_{\rho\sigma}\frac{-1}{96}(1+\frac{31}{12}\varepsilon)H^{2\varepsilon}Aa^{-4+\varepsilon}(\tau)\times
\big(a^2(\tau)\frac{\partial}{\partial a(\tau)}\big)^4
\Big\{\frac{\pi^{-\frac{\varepsilon}{2}}\mu^{-\varepsilon}}{\varepsilon\Gamma(1-\frac{\varepsilon}{2})}\log a(\tau)\notag\\
&+\log a(\tau)\log\frac{2\mu}{H}-\log^2a(\tau)
-a^2(\tau)\frac{\partial}{\partial a(\tau)}\int^{a(\tau)}_1da(\tau')\sum^\infty_{n=1}\frac{a^{n-2}(\tau')}{na^n(\tau)}\log a(\tau')\Big\}\notag\\
\simeq&\ 4i\pi^2\log a(\tau)\eta_{\rho\sigma}\cdot 0. \notag
\end{align}
Here we extract the terms which are proportional to $\log a(\tau)$. 
The integrals containing $F_{\rho\sigma}^2$ and $F_{\rho\sigma}^3$ are calculated analogously. 

Next, we calculate the integral containing $F_{\rho\sigma}^5$. 
\begin{equation}\begin{split}
&\ H^4A\int d^{4-\varepsilon}x'\ a^{4-\varepsilon}(\tau')\log a(\tau')F^5_{\rho\sigma}\\
=&\ H^{-4+2\varepsilon}Aa^{-5+\varepsilon}(\tau)
\int d^{4-\varepsilon}x'\ a(\tau')\log a(\tau')
\times 32(1-\frac{3}{4}\varepsilon)\big[\frac{\Delta x_\rho\Delta x_\sigma}{\Delta x^{10-2\varepsilon}_{++}}-\frac{\Delta x_\rho\Delta x_\sigma}{\Delta x^{10-2\varepsilon}_{+-}}\big]. 
\end{split}\label{A10}\end{equation}
The integrand is written as follows
\begin{equation}
\frac{\Delta x_\rho\Delta x_\sigma}{\Delta x^{10-2\varepsilon}}=\frac{-1}{2^5(4-\varepsilon)(3-\varepsilon)(2-\varepsilon)(1-\varepsilon)(1-\frac{\varepsilon}{2})}
\Big\{\partial_\rho\partial_\sigma+\frac{\eta_{\rho\sigma}\partial^2}{4-\varepsilon}\Big\}\partial^2\frac{\partial^2}{\varepsilon}\frac{1}{\Delta x^{2-2\varepsilon}}. 
\label{A11}\end{equation}
Note that the integral of the first term in (\ref{A11}) is proportional to $\delta_\rho^{\ 0}\delta_\sigma^{\ 0}$. 
So we neglect the first term in (\ref{A11}). 
The integral containing $F_{\rho\sigma}^5$ is identified as follows 
\begin{align}
&\ H^4A\int d^{4-\varepsilon}x'\ a^{4-\varepsilon}(\tau')\log a(\tau')F^5_{\rho\sigma}\label{A12}\\
\to&\ \eta_{\rho\sigma}\frac{-1}{96}(1+\frac{25}{12}\varepsilon)H^{-4+2\varepsilon}Aa^{-5+\varepsilon}(\tau)
\partial^4_0\int d^{4-\varepsilon}x'\ a(\tau')\log a(\tau')
\frac{\partial^2}{\varepsilon}\big[\frac{1}{\Delta x^{2-2\varepsilon}_{++}}-\frac{1}{\Delta x^{2-2\varepsilon}_{+-}}\big]. \notag
\end{align}
The subsequent processes are carried out just analogously with the integral containing $F_{\rho\sigma}^1$. 
The integrals containing $F_{\rho\sigma}^6$ and $F_{\rho\sigma}^7$ are calculated in a similar fashion. 

\subsection{Integrals containing $F_{\rho\sigma}^4$ and $F_{\rho\sigma}^8$}

We need a special consideration  when we evaluate integrals containing $F_{\rho\sigma}^4$ and $F_{\rho\sigma}^8$. 
These integrals consist of the two parts,  one part containing $1/\Delta x^{2n-2\varepsilon}$ 
and the other part containing $1/\Delta x^{2n-\varepsilon}$. 
Specifically the integral containing $F_{\rho\sigma}^4+F_{\rho\sigma}^8$ is written as follows 
\begin{align}
&\ H^4A\int d^{4-\varepsilon}x'a^{4-\varepsilon}(\tau')\log a(\tau')(F^4_{\rho\sigma}+F^8_{\rho\sigma})\label{A13}\\
=&\ \frac{\Gamma(4-\frac{\varepsilon}{2})}{4\Gamma(2-\frac{\varepsilon}{2})}H^{2\varepsilon}Aa^{-2+\varepsilon}(\tau)
\int d^{4-\varepsilon}x'a^2(\tau')\log a(\tau')
\Big\{\frac{\eta_{\rho\sigma}}{\Delta x^{4-2\varepsilon}}-(4-\varepsilon)\frac{a(\tau')}{a(\tau)}\frac{\Delta x_\rho\Delta x_\sigma}{\Delta x^{6-2\varepsilon}}\Big\}\notag\\
&-\frac{\Gamma(4-\varepsilon)}{2\Gamma(3-\frac{\varepsilon}{2})\Gamma(2-\frac{\varepsilon}{2})}
4^\frac{\varepsilon}{2}H^{\varepsilon}Aa^{-2+\frac{\varepsilon}{2}}(\tau)
\int d^{4-\varepsilon}x'a^{2-\frac{\varepsilon}{2}}(\tau')\log a(\tau')
\Big\{\frac{\eta_{\rho\sigma}}{\Delta x^{4-\varepsilon}}\notag\\
&\hspace{25em}-(4-\varepsilon)\frac{a(\tau')}{a(\tau)}\frac{\Delta x_\rho\Delta x_\sigma}{\Delta x^{6-\varepsilon}}\Big\}. \notag
\end{align}
We can evaluate the part containing $1/\Delta x^{2n-2\varepsilon}$ by the procedure we just explained. 
\begin{align}
&\ \frac{\Gamma(4-\frac{\varepsilon}{2})}{4\Gamma(2-\frac{\varepsilon}{2})}H^{2\varepsilon}Aa^{-2+\varepsilon}(\tau)
\int d^{4-\varepsilon}x'a^2(\tau')\log a(\tau')
\Big\{\frac{\eta_{\rho\sigma}}{\Delta x^{4-2\varepsilon}}-(4-\varepsilon)\frac{a(\tau')}{a(\tau)}\frac{\Delta x_\rho\Delta x_\sigma}{\Delta x^{6-2\varepsilon}}\Big\}\notag\\
\simeq&\ 4i\pi^2\log a(\tau)\eta_{\rho\sigma}\cdot\frac{-3}{16}. 
\label{A14}\end{align}

When we calculate the part containing $1/\Delta x^{2n-\varepsilon}$, 
we should note that the UV divergences at $\Delta x\sim 0$ are not regularized in the following integrals
even if $\varepsilon>0$  
\begin{equation}
\int d^{4-\varepsilon}x'\ \frac{1}{\Delta x^{2n-\varepsilon}},\hspace{1em}n\geq 2. 
\label{A15}\end{equation}
So we have to combine the terms in the integral so that these ill-defined terms don't appear. 
Herein the part containing $1/\Delta x^{2n-\varepsilon}$ is rewritten as follows 
\begin{align}
&-\frac{\Gamma(4-\varepsilon)}{2\Gamma(3-\frac{\varepsilon}{2})\Gamma(2-\frac{\varepsilon}{2})}
4^\frac{\varepsilon}{2}H^{\varepsilon}Aa^{-2+\frac{\varepsilon}{2}}(\tau)
\int d^{4-\varepsilon}x'a^{2-\frac{\varepsilon}{2}}(\tau')\log a(\tau')
\Big\{\frac{\eta_{\rho\sigma}}{\Delta x^{4-\varepsilon}}\notag\\
&\hspace{22em}-(4-\varepsilon)\frac{a(\tau')}{a(\tau)}\frac{\Delta x_\rho\Delta x_\sigma}{\Delta x^{6-\varepsilon}}\Big\}\notag\\
=&-\frac{\Gamma(4-\varepsilon)}{2\Gamma(3-\frac{\varepsilon}{2})\Gamma(2-\frac{\varepsilon}{2})}
4^\frac{\varepsilon}{2}H^{\varepsilon}Aa^{-2+\frac{\varepsilon}{2}}(\tau)
\int d^{4-\varepsilon}x'a^{2-\frac{\varepsilon}{2}}(\tau')\log a(\tau')
\Big\{\frac{\eta_{\rho\sigma}}{\Delta x^{4-\varepsilon}}\notag\\
&\hspace{12em}-(4-\varepsilon)\frac{\Delta x_\rho\Delta x_\sigma}{\Delta x^{6-\varepsilon}}
+(4-\varepsilon)a(\tau')\frac{H\Delta\tau\Delta x_\rho\Delta x_\sigma}{\Delta x^{6-\varepsilon}}\Big\}. 
\label{A16}\end{align}
By the power counting, 
it is found that the term containing $\Delta\tau\Delta x_\rho\Delta x_\sigma/\Delta x^{6-\varepsilon}$ is not divergent. 
To evaluate this term, we use the following identity 
\begin{equation}\begin{split}
\frac{\Delta\tau\Delta x_\rho\Delta x_\sigma}{\Delta x^{6-\varepsilon}}
=\frac{-1}{(4-\varepsilon)(2-\varepsilon)}
\Big\{&-\eta_{\rho\sigma}\partial_0\frac{1}{\Delta x^{2-\varepsilon}}
+\partial_0\partial_\sigma\frac{\Delta x_\rho}{\Delta x^{2-\varepsilon}}\\
&+(\delta_\rho^{\ 0}\partial_\sigma+\delta_\sigma^{\ 0}\partial_\rho)\frac{1}{\Delta x^{2-\varepsilon}}\Big\}. 
\end{split}\label{A17}\end{equation}
It is found that the residual term has no UV divergence from the following identities  
\begin{equation}\begin{split}
&\frac{\eta_{\rho\sigma}}{\Delta x^{4-\varepsilon}_{++}}-(4-\varepsilon)\frac{\Delta x_\rho\Delta x_\sigma}{\Delta x^{6-\varepsilon}_{++}}
=\frac{-1}{2-\varepsilon}\left\{\partial_\rho\partial_\sigma\frac{1}{\Delta x^{2-\varepsilon}_{++}}
+\frac{2(2-\varepsilon)ie\delta(\Delta\tau)\delta_\rho^{\ 0}\delta_\sigma^{\ 0}}{\Delta x_{++}^{4-\varepsilon}}\right\},\\
&\frac{\eta_{\rho\sigma}}{\Delta x^{4-\varepsilon}_{+-}}-(4-\varepsilon)\frac{\Delta x_\rho\Delta x_\sigma}{\Delta x^{6-\varepsilon}_{+-}}
=\frac{-1}{2-\varepsilon}\partial_\rho\partial_\sigma\frac{1}{\Delta x^{2-\varepsilon}_{+-}}. 
\end{split}\label{A18}\end{equation}

By substituting (\ref{A17}) and (\ref{A18}) to (\ref{A16}) and extract the terms which are proportional to $\eta_{\rho\sigma}\log a(\tau)$, 
\begin{align}
&-\frac{\Gamma(4-\varepsilon)}{2\Gamma(3-\frac{\varepsilon}{2})\Gamma(2-\frac{\varepsilon}{2})}
4^\frac{\varepsilon}{2}H^{\varepsilon}Aa^{-2+\frac{\varepsilon}{2}}(\tau)
\int d^{4-\varepsilon}x'a^{2-\frac{\varepsilon}{2}}(\tau')\log a(\tau')
\Big\{\frac{\eta_{\rho\sigma}}{\Delta x^{4-\varepsilon}}\notag\\
&\hspace{12em}-(4-\varepsilon)\frac{\Delta x_\rho\Delta x_\sigma}{\Delta x^{6-\varepsilon}}
+(4-\varepsilon)a(\tau')\frac{H\Delta\tau\Delta x_\rho\Delta x_\sigma}{\Delta x^{6-\varepsilon}}\Big\}\notag\\
\to&-\eta_{\rho\sigma}\frac{3}{4}Ha^{-2}(\tau)\partial_0\int d^4x'a^3(\tau')\log a(\tau')
\times\big[\frac{1}{\Delta x^2_{++}}-\frac{1}{\Delta x^2_{+-}}\big]\notag\\
\simeq&\ 4i\pi^2\log a(\tau)\ \eta_{\rho\sigma}\cdot\frac{3}{8}. 
\label{A19}\end{align}
Here we have used the identity
\begin{equation}
\frac{1}{\Delta x^2}=\frac{1}{4}\partial^2\log(\mu^2\Delta x^2), 
\label{A20}\end{equation}
and (\ref{A7}). 
From (\ref{A13}), (\ref{A14}) and (\ref{A19}), we obtain
\begin{equation}\begin{split}
&\ H^4A\int d^{4-\varepsilon}x'a^{4-\varepsilon}(\tau')\log a(\tau')(F^4_{\rho\sigma}+F^8_{\rho\sigma})\\
\simeq&\ 4i\pi^2\log a(\tau)\ \eta_{\rho\sigma}\cdot\frac{3}{16}. 
\end{split}\label{A21}\end{equation}

\section{Two point function in the non-linear sigma model}
\setcounter{equation}{0}
The two point function at the 2 loop level is evaluated as follows
\begin{align}
&\ \langle\xi^a\xi^a\rangle|_{g^2}\label{B1}\\
=&\ \int d^Dx'\ a^D(\tau')
\ i\frac{g^2}{3}R\lim_{x''\to x'} \partial_\alpha'\partial_\beta'' G^{++}(x',x'')\notag\\
&\times g^{\alpha\beta}(\tau')
\big[G^{++}(x,x')G^{++}(x,x')-G^{+-}(x,x')G^{+-}(x,x')\big]\notag\\
&-\int d^Dx'\ a^D(\tau')
\ i\frac{g^2}{6}R\partial_\alpha' G^{++}(x',x')\notag\\
&\times g^{\alpha\beta}(\tau')
\partial_\beta' \big[G^{++}(x,x')G^{++}(x,x')-G^{+-}(x,x')G^{+-}(x,x')\big]\notag\\
&+\int d^Dx'\ a^D(\tau')
\big\{i\frac{g^2}{3}RG^{++}(x',x')-iN(\delta Z-\frac{\delta g^2}{g^2})\big\}\notag\\
&\times g^{\alpha\beta}(\tau')
\big[\partial_\alpha' G^{++}(x,x')\partial_\beta' G^{++}(x,x')
-\partial_\alpha' G^{+-}(x,x')\partial_\beta' G^{+-}(x,x')\big]. \notag
\end{align}
We extract the dominant terms which are proportional to $g^2\log^2 a(\tau)$. 
We find that only the following part of the propagator contributes to the dominant terms 
\begin{align}
G(x,x')\simeq-\frac{H^2}{8\pi^2}\log H^2\Delta x^2 . \notag
\end{align}
The contribution from each diagram is as follows 
\begin{align}
&\ \int d^Dx'\ a^D(\tau')
\ i\frac{g^2}{3}R\lim_{x''\to x'} \partial_\alpha'\partial_\beta'' G^{++}(x',x'')\label{B3}\\
&\times g^{\alpha\beta}(\tau')
\big[G^{++}(x,x')G^{++}(x,x')-G^{+-}(x,x')G^{+-}(x,x')\big]\notag\\
\simeq&-i\frac{g^2RH^8}{2^9\pi^6}\int d^4x'\ a^4(\tau') \big[\log^2(H^2\Delta x^2_{++})-\log^2(H^2\Delta x^2_{+-})\big]\notag\\
\simeq&-\frac{g^2RH^4}{2^5\cdot 3\pi^4}\log^2 a(\tau), \notag
\end{align}
\begin{align}
&-\int d^Dx'\ a^D(\tau')
\ i\frac{g^2}{6}R\partial_\alpha' G^{++}(x',x')\label{B4}\\
&\times g^{\alpha\beta}(\tau')
\partial_\beta' \big[G^{++}(x,x')G^{++}(x,x')-G^{+-}(x,x')G^{+-}(x,x')\big]\notag\\
\simeq&+i\frac{g^2RH^7}{2^9\cdot 3\pi^6}\int d^4x'\ a^3(\tau') \partial_0'\big[\log^2(H^2\Delta x^2_{++})-\log^2(H^2\Delta x^2_{+-})\big]\notag\\
\simeq&-\frac{g^2RH^4}{2^5\cdot 3\pi^4}\log^2 a(\tau), \notag
\end{align}
\begin{align}
&\ \int d^Dx'\ a^D(\tau')
\big\{i\frac{g^2}{3}RG^{++}(x',x')-iN(\delta Z-\frac{\delta g^2}{g^2})\big\}\label{B5}\\
&\times g^{\alpha\beta}(\tau')
\big[\partial_\alpha' G^{++}(x,x')\partial_\beta' G^{++}(x,x')
-\partial_\alpha' G^{+-}(x,x')\partial_\beta' G^{+-}(x,x')\big]\notag\\
\simeq&+i\frac{g^2RH^6}{2^6\cdot 3\pi^6}\int d^4x'\ a^2(\tau')\log a(\tau') \big[\frac{1}{\Delta x^2_{++}}-\frac{1}{\Delta x^2_{+-}}\big]\notag\\
\simeq&+\frac{g^2RH^4}{2^5\cdot 3\pi^4}\log^2 a(\tau). \notag
\end{align}
Note that the power of the leading IR logarithm is not equal to 
the number of propagators left intact by the derivatives. 
Time derivative operator $\partial_0'$ increases the power of scale factor $a(\tau')$ 
while the inverse of metric $g^{\mu\nu}(\tau')$ decreases it. 
From (\ref{B1}), (\ref{B3}), (\ref{B4}) and (\ref{B5}), 
\begin{align}
&\ \langle\xi^a\xi^a\rangle|_{g^2}\simeq-\frac{g^2RH^4}{2^5\cdot 3\pi^4}\log^2 a(\tau). 
\end{align}


\end{document}